\documentclass[12pt,a4paper]{article}
\usepackage{eurosym}
\usepackage{amsfonts}
\usepackage{amssymb}
\usepackage{amsmath}
\usepackage{amsthm}
\usepackage{amstext}
\usepackage[dvips]{graphicx}
\usepackage{graphics}
\usepackage{ulem}
\usepackage[usenames,dvipsnames]{xcolor}
\usepackage[T1]{fontenc}
\usepackage{epsfig}
\usepackage{soul}
\usepackage{subfig}
\usepackage[]{natbib}
\usepackage[]{enumerate}
\usepackage{booktabs}
\usepackage{rotating}
\usepackage{caption}
\usepackage{multirow}

\usepackage{tikz}
\usetikzlibrary{calc,decorations.pathreplacing}
\usetikzlibrary{arrows}
\usetikzlibrary{patterns}
\usetikzlibrary{fadings}

\emergencystretch=\maxdimen
\newtheoremstyle%
 {redthm}%
 {}{}%
 {\color{black}\itshape}
 {}%
 {\color{OrangeRed}\bfseries}%
 {\color{OrangeRed}.}%
 { }{}

\newtheoremstyle%
 {orangethm}%
 {}{}%
 {\color{black}\itshape}
 {}%
 {\color{RedOrange}\bfseries}%
 {\color{RedOrange}.}%
 { }{}

\theoremstyle{redthm}

\theoremstyle{orangethm}

\theoremstyle{orangethm}

\begin{document}

\title{
Measuring individual contributions to team success in professional road cycling\thanks{I would like to thank Juan D. Moreno-Ternero and Ricardo Mart\'{i}nez for their helpful comments. The usual disclaimer applies.}}

\author{\textbf{Aitor Calo-Blanco\thanks{Corresponding author at: Departamento de Econom\'{i}a, Universidade da Coru\~{n}a, Campus de Elvi\~{n}a, 15071 A Coru\~{n}a, Spain. E-mail: aitor.calo@udc.es. ORCID ID: 0000-0003-3957-4741.}} \\ Universidade da Coru\~{n}a}


\maketitle

\begin{abstract}

Assessing individual contributions within teams remains a fundamental challenge. This issue is particularly important in professional road cycling, where personal success hinges on a combination of individual talent and collective effort. Current points-based rankings disproportionately reward team leaders, undervaluing those who contribute through collaborative roles. To address this issue, we propose a flexible framework for collaborative contribution within professional cycling teams that combines individual performance, race participation, and a team-based credit-allocation system. Using data from the 2023 season, we show that our framework identifies meaningful differences in rider contribution that existing ranking systems fail to capture. We then select a parameter configuration for our framework that closely aligns with rider assessments from Pro Cycling Manager, assigning equal weights to individual performance, race participation, and collaborative contribution. Finally, we use this configuration to examine which rider characteristics are associated with gains or losses under collaborative evaluation.

\vspace{0.15cm}


\vspace{0.05cm}


\end{abstract}


\section{Introduction}
\label{sect_intro}

Quantifying individual contributions within collaborative production systems is a fundamental problem of credit assignment. Since observable outcomes often reflect collective rather than individual effort, evaluating contributors has become an important topic in sports analytics. Professional road cycling provides a particularly suitable setting in which to study this problem because it is characterized by a distinctive interplay between individual performance and team effort. Although victories are achieved by a single rider, success crucially depends on domestique riders, who sacrifice personal success to support group performance \citep[see][]{Torgler_07_JSE, Rodriguez_14_IJSF}. Consequently, evaluating riders requires a framework that can allocate collective output among collaborators while recognizing individual achievement and collaborative contribution.

\begin{figure}[t]
  \centering
    \subfloat[Between and within teams]{\label{fig_barh}$
        \includegraphics[height=55mm]{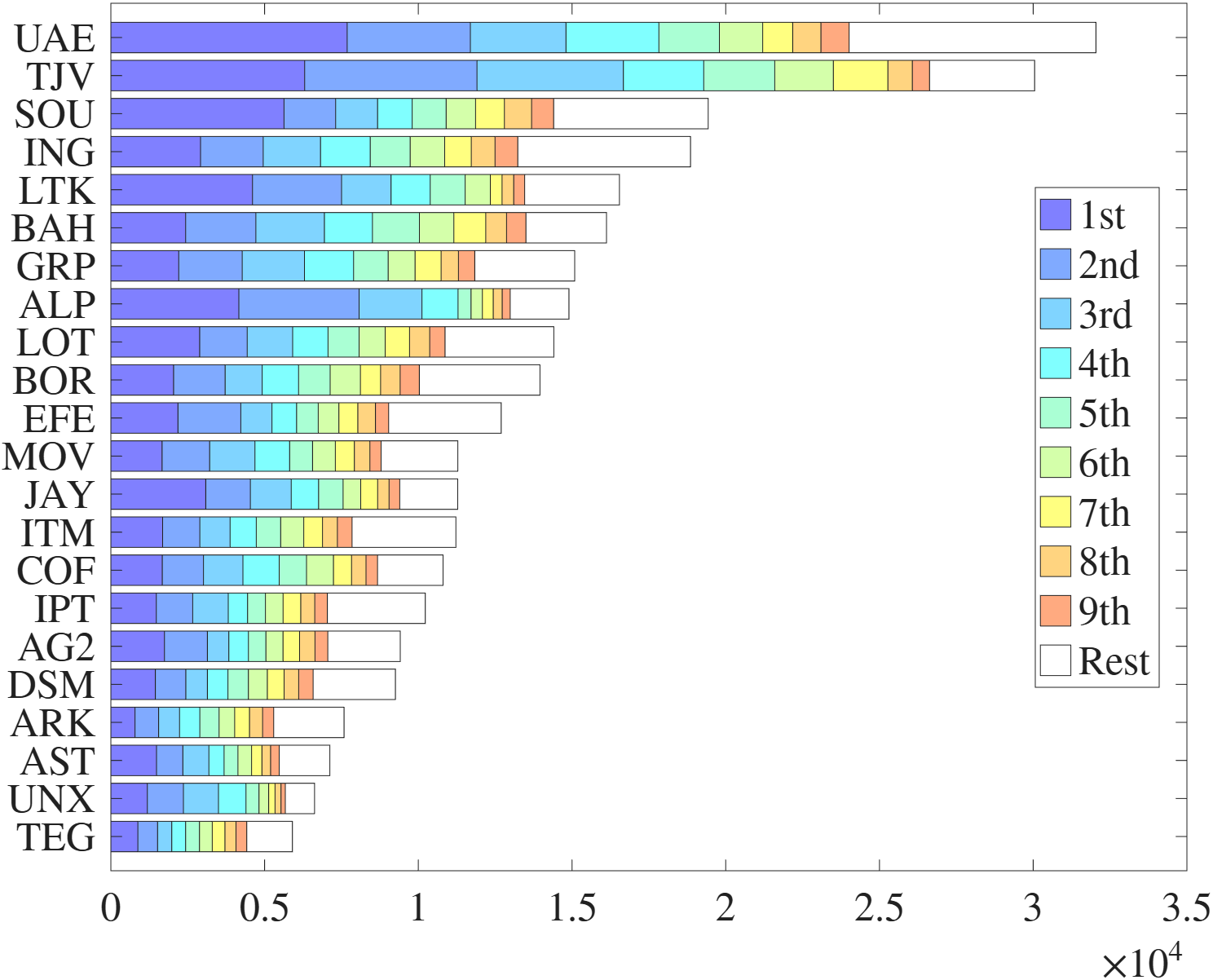}
      $}
      \hspace{0.10cm}
      \subfloat[Top five riders highlighted in purple. Riders ranked 6th to 10th in cyan]{\label{fig_rasc}$
    \includegraphics[height=55mm]{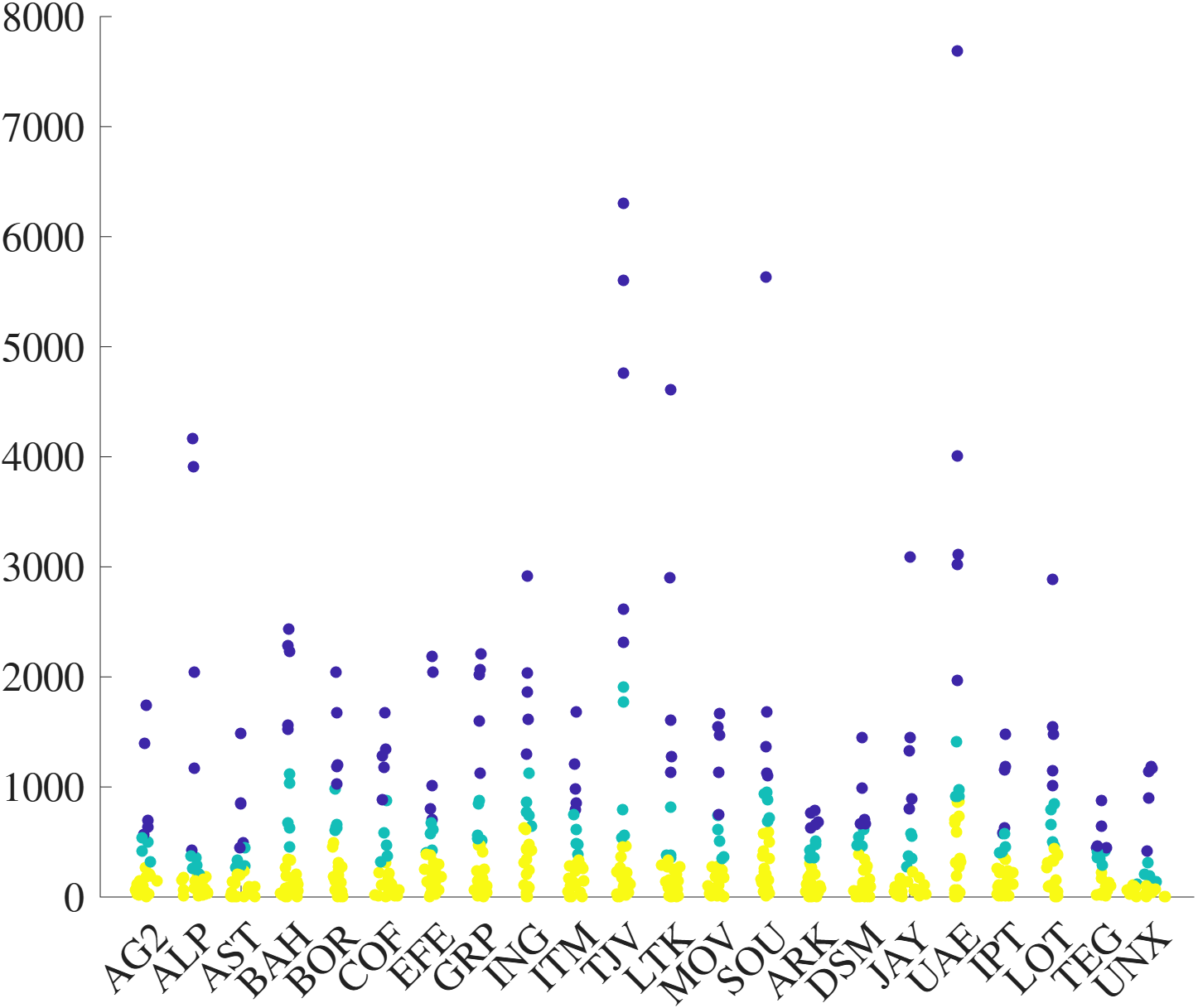}
    $}
    \caption{Distribution of UCI points among the top 22 teams in the 2023 season}
    \label{fig_distribution_UCI}
  \end{figure}

The Union Cycliste Internationale (UCI), the world governing body for professional cycling, uses a points-based system to rank teams and riders. This ranking is important because it determines team eligibility for the most prestigious events and affects the profitability of professional teams. Riders also place high value on UCI points, as they may influence their selection for races and future contracts. However, the UCI points system rewards individual outcomes, thus failing to identify what teams value internally. As illustrated in Figure \ref{fig_distribution_UCI}, during the 2023 season the top nine scorers in each of the 22 best teams (listed in Table \ref{tab_data_A}), representing 30\% of their rosters, accumulated more than three quarters of the total points for these teams. This high degree of concentration makes the UCI points-based system an imperfect measure of collaborative contribution.

To address this issue, we propose the Rider Contribution Score (RCS), a flexible framework for allocating team output among riders. The RCS consists of a convex combination of three different performance indicators: UCI points, a proportional redistribution of these points based on race participation, and a version of the CoScore formula introduced by \cite{Flores-Szwagrzak_al_20_MS}. CoScore defines each team member's value as the sum of the credits assigned to them across all projects in which they have participated. Thus, it evaluates contributors based not only on their own achievements but also on the performance of their teammates.

The RCS is a useful tool for sports managers and data analysts because it gives them discretion to balance the relative importance of the three underlying components. First, incorporating UCI points ensures that the method aligns with the main factor determining professional rankings. However, moving beyond UCI points helps riders and managers to address the inherent conflict between individual achievement and collective success. Second, the use of a participation-weighted redistribution of points compensates riders who make sacrifices for the team. Including the other two indicators prevents the overvaluation of riders based solely on their participation in races, regardless of their actual performance. Finally, the use of CoScore allows us to credit riders for points earned by their teams in the races in which they participate. Combining CoScore with the other two indicators mitigates its potential negative effect of undervaluing valuable riders who consistently race with team leaders.

We apply our rider contribution framework to all professional riders in the top 22 teams during the 2023 professional road cycling season. The analysis reveals that the RCS identifies meaningful differences between riders that current ranking systems are not able to capture. We then use an external benchmark to select a parameter configuration for the RCS that aligns with expert assessments of riders. To conduct this analysis, we use data from the popular simulation game Pro Cycling Manager. This game provides rider ratings developed by data analysts using historical results and community feedback. Finally, we use this configuration to examine which rider characteristics are associated with gains or losses under collaborative evaluation. Regression analyses show that the benchmark configuration systematically redistributes value toward lower-ranked and more team-oriented riders. Race participation and team strength are positively associated with gains under the benchmark configuration. Furthermore, high-scoring riders tend to experience larger reductions in their assigned value, with the effects varying systematically across rider specializations.

Teamwork and individual performance have been studied in various sports, including football \citep{Saebo_al_19_JSA}, basketball \citep{Cooper_al_09_EJOR, Engelmann_17}, baseball \citep[][]{Barnes_al_24_AOR}, and volleyball \citep{Hass_al_19_JSA, Lopez_Serrano_al_24_IJPAS}. In professional road cycling, most studies have focused on assessing or predicting rider performance and value \citep[e.g.,][]{Cherchye_al_06_JSE, Van_Bulck_al_23_AOR, Janssens_23_AOR}. However, because these studies primarily rely on individual race results, they do not directly address the potential undervaluation of domestiques. Our assessment of collaborative contribution is related to the plus-minus statistic, which captures aspects of team performance that do not appear in traditional individual statistics. However, plus-minus models typically rely on detailed microdata and can be difficult to interpret for managerial decision-making. In contrast, the RCS is based on publicly available cycling data and provides a transparent framework that managers and analysts can readily implement and interpret.


\section{Materials and methods}
\label{sect_material_method}


\subsection{The measure of rider contribution}
\label{subsect_RCS}

A road cycling season is a triple $(R,T,N)$, where $R$ denotes the set of races (including their stages and points), $T$ the set of teams, and $N$ the set of riders. The group of riders enrolled in team $t\in T$ is denoted by $N(t)$, and the subset of them who participate in race $r\in R$ by $N(r,t)$. For any rider $i\in N(t)$, the set of races in which this rider participates is denoted by $R_i(t)$. Let $R(t)$ represent the races in which team $t$ takes part with at least one rider. $p_{i}(r,t)\in\mathbb{R}_{+}$ denotes the UCI points scored by rider $i\in N(t)$ in race $r\in R_{i}(t)$, and $P_{i}(t)=\sum_{r\in R_{i}(t)}p_{i}(r,t)$ denotes the total points rider $i$ scores during the season. For any team $t\in T$, let $p(r,t)=\sum_{i\in N(r,t)}p_{i}(r,t)$ be the points this team earns in race $r\in R$, and let $P(t)=\sum_{r\in R(t)}p(r,t)$ be the total points it earns during the entire season.

Our analysis of collaborative contribution within road cycling aims to allocate a team's total UCI points among its riders. To do so, we define the Rider Contribution Score (RCS), which is constructed based on three different elements. Each element provides a distinct benchmark for allocating team points among riders. First, we consider the actual distribution of points earned by each rider throughout the season. Second, to capture each rider's exposure to collective success, we consider a redistribution of the team's total points weighted by individual participation. Road cycling includes both one-day races and stage races, which means that the level of participation required to score UCI points differs between races. Therefore, for any race $r\in R_{i}(t)$, let $d_{i}(r,t)\in\mathbb{R}_{+}$ be the number of race days started by rider $i\in N(t)$. The total number of race days started by team $t\in T$ in race $r\in R(t)$ is denoted by $d(r,t)=\sum_{i\in N(r,t)}d_{i}(r,t)$.

The third component of the RCS is a version of the CoScore formula introduced by \cite{Flores-Szwagrzak_al_20_MS}. This formula provides a natural starting point for measuring rider contribution because it defines a team member's value as the sum of the credits assigned to them across all projects in which they have participated. Since each rider's value depends partly on the value of their teammates, individual contributions must be determined simultaneously as the solution to a fixed-point problem rather than independently. As shown by \cite{Flores-Szwagrzak_al_20_MS}, CoScore is consistent with respect to three properties: the removal of departing agents, the division of collaborative projects into subprojects, and the aggregation of team members into subteams. Consequently, the measure behaves predictably under natural changes in team structure.

\begin{figure}
  \centering
    \includegraphics[height=60mm]{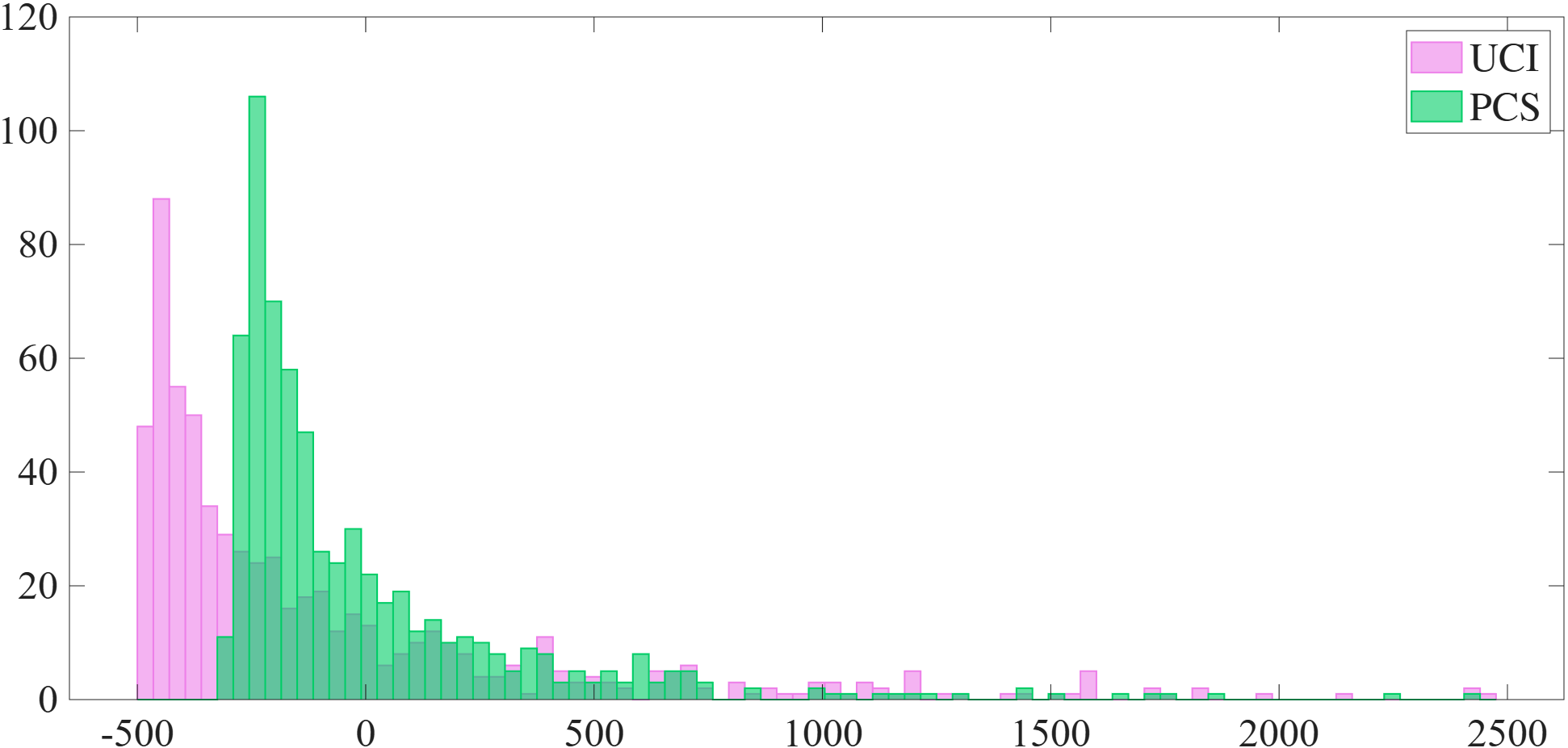}
  \caption{Distribution of UCI and PCS points, with respect to the mean, during the 2023 season (limited to 2500 points)}
  \label{fig_pcs}
  \end{figure}

Because cyclists assume roles of differing strategic importance, they should not be treated as equally important contributors. To account for this, we incorporate race-specific weights based on PCS points. ProCyclingStats (PCS) is a database that compiles race results and provides a consistent points-based assessment that is widely used in the cycling community. Although PCS and UCI points are highly correlated, PCS has a narrower range between its highest and lowest values. This is because PCS awards points to riders who finish in top-tier races and limits the points awarded to the highest-ranked finishers. Figure \ref{fig_pcs} compares the empirical distributions of PCS and UCI points for the 2023 season. Our weights are defined as follows:\footnote{We define PCS points similarly to how we defined UCI points. The number of PCS points obtained by rider $i$ enrolled in team $t$ during race $r$ is denoted as $\widehat{p}_{i}(r,t)$.}

  \begin{equation}
    \omega_{i}(r,t)=\frac{\widehat{p}_{i}(r,t)}{\widehat{p}(r,t)}, \ \textnormal{for all} \ i\in N. 
  \end{equation}

\vspace{0.25cm}

In the few races for which a team earns UCI points but not PCS points, we assign equal weights to all participating riders. By construction, the weights add up to one across riders participating in each race.

The RCS evaluates rider contribution through a weighted aggregation of three measures of cycling performance: individual achievement, participation contribution, and collaborative contribution. First, the formula identifies individual achievement with UCI points, assigning this measure a weight $\alpha\in[0,1]$. Second, the participation measure, which is weighted with the parameter $\beta\in[0,1]$, is defined by the proportional redistribution of the team's total UCI points based on race participation. For $\alpha,\beta\in[0,1]$ satisfying $0<\alpha+\beta\leq1$, the RCS defines a convex combination of the three measures of cycling performance in which the weighted version of CoScore accounts for the individual collaborative contribution.

A points allocation for a cycling season $s=(R,T,N)$ is a vector $x(s)\in\mathbb{R}^{N}_{+}$ such that $\sum_{i\in N}x_{i}(s)=\sum_{t\in T}P(t)$. To select a points allocation, the RCS combines the three previous components into a single fixed-point allocation rule. More precisely, for all $i\in N$,

  \begin{equation}
  \begin{split}
  x^{R}_{i}(s)&=\alpha P_{i}(t) + \beta\sum_{r\in R_{i}(t)}\left[\frac{d_{i}(r,t)}{d(r,t)}\right]p(r,t)
  \\
  & + (1-\alpha-\beta)\sum_{r\in R_{i}(t)} \left[ \frac{d_{i}(r,t) \omega_{i}(r,t)x^{R}_{i}(s)}{\sum\limits_{j\in N(r,t)}d_{j}(r,t) \omega_{j}(r,t)x^{R}_{j}(s)}
    \right]p(r,t).
  \end{split}
  \label{eq_dw_CoS}
  \end{equation}

\vspace{0.25cm}

Note that $x^{R}_{i}(s)$ appears on both sides of equation \eqref{eq_dw_CoS}. This reflects the CoScore component of our measure of rider contribution, which describes the credit assigned to rider $i$ for shared projects. Under the parameter restriction $0<\alpha+\beta\leq1$, the fixed point exists and is unique, as shown by \cite{Flores-Szwagrzak_al_20_MS}. In addition, the iterative procedure converges rapidly to this fixed point.

The limitations of relying on a single performance measure are illustrated by the following example. Three riders from the same team compete in four different races. Riders 2 and 3 are the team leaders, while rider 1 is a strong cyclist who sacrifices personal success to assist the team leaders. In the first race, only riders 1 and 2 compete for the team, scoring 50 and 950 points, respectively. In the second event, rider 1 competes with rider 3, and they score 100 and 900 points, respectively. In the third race, rider 1 obtains 200 points competing alone. All three riders participate in the fourth race, scoring 200, 1200, and 600 points, respectively. Together, they accumulate a total of 4200 points. These results are summarized in Figure \ref{fig_ex_uci}.

As rider 1 participates in all races, the participation-weighted redistribution of points clearly favors this rider, who receives $44\%$ of the total points (see Figure \ref{fig_ex_part}, where actual points are shown by the dotted lines). No distinction is made between riders 2 and 3, since the total number of points earned by the team in the races in which they participate is the same. Assuming $\alpha+\beta=0.1$, the unweighted version of CoScore assigns $84\%$ of the team's total points to rider 1 (Figure \ref{fig_ex_cosc_1}). Since this measure is unweighted, rider 1's participation in every race leads CoScore to attribute most of the team's success to this rider.

However, the weighted version of CoScore, also assuming $\alpha+\beta=0.1$, yields the opposite result, reducing rider 1's value to 286.55 points, less than $7\%$ of the team's total points (Figure \ref{fig_ex_cosc_2}). Because rider 1 consistently scores fewer points than both teammates, the weighted CoScore assigns this rider a collaborative contribution only slightly above the value generated when riding alone. Therefore, using only CoScore may either overvalue riders who participate in many races or undervalue relevant scorers who support team leaders. Consequently, combining complementary indicators provides a more balanced measure of collaborative contribution in road cycling.

  \begin{figure}[t] 
  \centering
    \subfloat[Points]{\label{fig_ex_uci}$ 
    \begin{tikzpicture}[x=0.75cm,y=0.5cm,z=0.30cm,>=stealth, scale=0.92, every node/.style={transform shape}]
              \filldraw[red!35,fill=red!35,opacity=0.40] (3.00,0.00) -- (3.00,1.00) -- (1.50,1.00) -- (1.50,0.00) -- (3.00,0.00);
              \filldraw[green!35,fill=green!35,opacity=0.40] (0.00,0.00) -- (0.00,5.00) -- (2.00,5.00) -- (2.00,1.00) -- (1.50,1.00) -- (1.50,0.00) -- (0.00,0.00);
              \filldraw[blue!35,fill=blue!25,opacity=0.40] (4.00,0.00) -- (4.00,5.00) -- (2.00,5.00) -- (2.00,1.00) -- (3.00,1.00) -- (3.00,0.00) -- (4.00,0.00);
              \filldraw[red!35,fill=red!35,opacity=0.40] (0.00,0.00) -- (4.00,0.00) -- (4.00,-0.50) -- (0.00,-0.50) -- (0.00,0.00);
              \filldraw[red!35,fill=red!35,opacity=0.40] (1.00,-0.50) -- (1.00,-1.50) -- (3.00,-1.50) -- (3.00,-0.50);
              \filldraw[green!35,fill=green!35,opacity=0.40] (0.00,-0.50) -- (1.00,-0.50) -- (1.00,-1.50) -- (2.75,-1.50) -- (2.75,-5.50) -- (0.00,-5.50) -- (0.00,-0.50);
              \filldraw[blue!35,fill=blue!25,opacity=0.40] (4.00,-0.50) -- (3.00,-0.50) -- (3.00,-1.50) -- (2.75,-1.50) -- (2.75,-5.50) -- (4.00,-5.50) -- (4.00,-0.50);
              \draw[line width=0.25mm,gray,dotted] (1.50,0.00) -- (1.50,1.00) -- (3.00,1.00) -- (3.00,0.00);
              \draw[line width=0.25mm,gray,dotted] (3.00,-0.50) -- (3.00,-1.50) -- (1.00,-1.50) -- (1.00,-0.50);
              \draw[line width=0.25mm,gray,dotted] (2.75,-1.50) -- (2.75,-5.50);
              \draw[line width=0.20mm,black] (0.00,5.00) -- (4.00,5.00) -- (4.00,-5.50) -- (0.00,-5.50) -- (0.00,5.00);
              \draw[line width=0.20mm,black] (0.00,0.00) -- (4.00,0.00);
              \draw[line width=0.20mm,black] (0.00,-0.50) -- (4.00,-0.50);
              \draw[line width=0.20mm,black] (2.00,5.00) -- (2.00,0.00);
              \node[align=center,red] at (1.75,0.50) (ori) {$\textbf{1}$};
              \node[align=center,red] at (2.50,0.50) (ori) {$\textbf{1}$};
              \node[align=center,red] at (0.50,-0.25) (ori) {$\textbf{\scriptsize{1}}$};
              \node[align=center,PineGreen] at (1.00,2.85) (ori) {$\textbf{2}$};
              \node[align=center,blue] at (3.00,2.85) (ori) {$\textbf{3}$};
              \node[align=center,red] at (2.00,-1.00) (ori) {$\textbf{1}$};
              \node[align=center,PineGreen] at (1.375,-3.25) (ori) {$\textbf{2}$};
              \node[align=center,blue] at (3.375,-3.25) (ori) {$\textbf{3}$};
    \end{tikzpicture}
      $}
      \hspace{0.45cm}
    \subfloat[Participation]{\label{fig_ex_part}$ 
    \begin{tikzpicture}[x=0.75cm,y=0.5cm,z=0.30cm,>=stealth, scale=0.92, every node/.style={transform shape}]
              \filldraw[red!35,fill=red!35,opacity=0.40] (0.00,2.06) -- (4.00,2.06) -- (4.00,-2.56) -- (0.00,-2.56) -- (0.00,2.06);
              \filldraw[green!35,fill=green!35,opacity=0.40] (0.00,5.00) -- (2.00,5.00) -- (2.00,2.06) -- (0.00,2.06);
              \filldraw[blue!35,fill=blue!25,opacity=0.40] (2.00,2.06) -- (2.00,5.00) -- (4.00,5.00) -- (4.00,2.06);
              \filldraw[green!35,fill=green!35,opacity=0.40] (0.00,-2.56) -- (2.00,-2.56) -- (2.00,-5.50) -- (0.00,-5.50) -- (0.00,-2.56);
              \filldraw[blue!35,fill=blue!25,opacity=0.40] (2.00,-2.56) -- (2.00,-5.50) -- (4.00,-5.50) -- (4.00,-2.56);
              \draw[line width=0.25mm,gray,dotted] (0.00,0.00) -- (1.50,0.00) -- (1.50,1.00) -- (3.00,1.00) -- (3.00,0.00) --(4.00,0.00);
              \draw[line width=0.25mm,gray,dotted] (4.00,-0.50) -- (3.00,-0.50) -- (3.00,-1.50) -- (1.00,-1.50) -- (1.00,-0.50) -- (0.00,-0.50);
              \draw[line width=0.25mm,gray,dotted] (2.00,5.00) -- (2.00,1.00);
              \draw[line width=0.25mm,gray,dotted] (2.75,-1.50) -- (2.75,-5.50);
              \draw[line width=0.20mm,black] (0.00,5.00) -- (4.00,5.00) -- (4.00,-5.50) -- (0.00,-5.50) -- (0.00,5.00);
               \node[align=center,red] at (2.00,-0.25) (ori) {$\textbf{1}$};
               \node[align=center,PineGreen] at (1.00,3.40) (ori) {$\textbf{2}$};
               \node[align=center,blue] at (3.00,3.40) (ori) {$\textbf{3}$};
               \node[align=center,PineGreen] at (1.00,-4.15) (ori) {$\textbf{2}$};
               \node[align=center,blue] at (3.00,-4.15) (ori) {$\textbf{3}$};
    \end{tikzpicture}
      $}
      \hspace{0.45cm}
    \subfloat[CoScore]{\label{fig_ex_cosc_1}$ 
    \begin{tikzpicture}[x=0.75cm,y=0.5cm,z=0.30cm,>=stealth, scale=0.92, every node/.style={transform shape}]
              \filldraw[red!35,fill=red!35,opacity=0.40] (0.00,4.16) -- (4.00,4.16) -- (4.00,-4.66) -- (0.00,-4.66) -- (0.00,4.16);
              \filldraw[green!35,fill=green!35,opacity=0.40] (0.00,5.00) -- (2.00,5.00) -- (2.00,4.16) -- (0.00,4.16);
              \filldraw[blue!35,fill=blue!25,opacity=0.40] (2.00,4.16) -- (2.00,5.00) -- (4.00,5.00) -- (4.00,4.16);
              \filldraw[green!35,fill=green!35,opacity=0.40] (0.00,-4.66) -- (2.00,-4.66) -- (2.00,-5.50) -- (0.00,-5.50);
              \filldraw[blue!35,fill=blue!25,opacity=0.40] (2.00,-4.66) -- (2.00,-5.50) -- (4.00,-5.50) -- (4.00,-4.66);
              \draw[line width=0.25mm,gray,dotted] (0.00,0.00) -- (1.50,0.00) -- (1.50,1.00) -- (3.00,1.00) -- (3.00,0.00) --(4.00,0.00);
              \draw[line width=0.25mm,gray,dotted] (4.00,-0.50) -- (3.00,-0.50) -- (3.00,-1.50) -- (1.00,-1.50) -- (1.00,-0.50) -- (0.00,-0.50);
              \draw[line width=0.25mm,gray,dotted] (2.00,5.00) -- (2.00,1.00);
              \draw[line width=0.25mm,gray,dotted] (2.75,-1.50) -- (2.75,-5.50);
              \draw[line width=0.20mm,black] (0.00,5.00) -- (4.00,5.00) -- (4.00,-5.50) -- (0.00,-5.50) -- (0.00,5.00);
              \node[align=center,red] at (2.00,-0.25) (ori) {$\textbf{1}$};
              \node[align=center,PineGreen] at (1.00,4.55) (ori) {$\textbf{2}$};
              \node[align=center,blue] at (3.00,4.55) (ori) {$\textbf{3}$};
              \node[align=center,PineGreen] at (1.00,-5.00) (ori) {$\textbf{2}$};
              \node[align=center,blue] at (3.00,-5.00) (ori) {$\textbf{3}$};
    \end{tikzpicture}
      $}
      \hspace{0.45cm}
    \subfloat[$\omega$CoScore]{\label{fig_ex_cosc_2}$ 
    \begin{tikzpicture}[x=0.75cm,y=0.5cm,z=0.30cm,>=stealth, scale=0.92, every node/.style={transform shape}]
              \filldraw[red!35,fill=red!35,opacity=0.40] (0.00,0.00) -- (2.53,0.00) -- (2.53,1.00) -- (3.00,1.00) -- (3.00,0.00) -- (4.00,0.00)  -- (4.00,-0.50)   -- (3.00,-0.50) -- (3.00,-1.50) -- (2.53,-1.50) -- (2.53,-0.50) -- (0.00,-0.50);
              \filldraw[green!35,fill=green!35,opacity=0.40] (0.00,5.00) -- (2.09,5.00) -- (2.09,1.00) -- (2.53,1.00) -- (2.53,0.00) -- (0.00,0.00);
              \filldraw[blue!35,fill=blue!25,opacity=0.40] (2.09,5.00) -- (4.00,5.00) -- (4.00,0.00) -- (3.00,0.00) -- (3.00,1.00) -- (2.09,1.00);
              \filldraw[green!35,fill=green!35,opacity=0.40] (0.00,-0.50) -- (2.53,-0.50) -- (2.53,-1.50) -- (2.84,-1.50) -- (2.84,-5.50) -- (0.00,-5.50);
              \filldraw[blue!35,fill=blue!25,opacity=0.40] (2.84,-1.50) -- (3.00,-1.50) -- (3.00,-0.50) -- (4.00,-0.50) -- (4.00,-5.50) -- (2.84,-5.50);
              \draw[line width=0.25mm, gray,dotted] (0.00,0.00) -- (1.50,0.00) -- (1.50,1.00) -- (3.00,1.00) -- (3.00,0.00) --(4.00,0.00);
              \draw[line width=0.25mm, gray,dotted] (4.00,-0.50) -- (3.00,-0.50) -- (3.00,-1.50) -- (1.00,-1.50) -- (1.00,-0.50) -- (0.00,-0.50);
              \draw[line width=0.25mm, gray,dotted] (2.00,5.00) -- (2.00,1.00);
              \draw[line width=0.25mm, gray,dotted] (2.75,-1.50) -- (2.75,-5.50);
              \draw[line width=0.20mm, black] (0.00,5.00) -- (4.00,5.00) -- (4.00,-5.50) -- (0.00,-5.50) -- (0.00,5.00);
              \node[align=center, red] at (2.80,-0.25) (ori) {$\textbf{1}$};
              \node[align=center, PineGreen] at (1.05,2.85) (ori) {$\textbf{2}$};
              \node[align=center, blue] at (3.05,2.85) (ori) {$\textbf{3}$};
              \node[align=center, PineGreen] at (1.38,-3.25) (ori) {$\textbf{2}$};
              \node[align=center, blue] at (3.38,-3.25) (ori) {$\textbf{3}$};
    \end{tikzpicture}
      $}
    \caption{Rider contribution with different measures}
    \label{fig_example}
  \end{figure}

Finally, our framework is modular: analysts and managers may incorporate additional measures of cycling performance if they consider them relevant to the evaluation. 


\subsection{The UCI points ranking system}
\label{subsect_regulation}

We focus on the 2023 professional road cycling season and its team structure. This season marks the start of a new three-year promotion-relegation cycle. We analyze elite male riders belonging to the top 22 teams in our sample -- the 18 WT teams and the top four ProTeams (the second tier) -- during the 2023 season. We include their results in all WT, ProSeries (x.Pro), and professional Continental (x.1) races. World and continental championships feature cyclists organized by national teams rather than sponsored squads. National championships feature cyclists in private teams of varying sizes. Therefore, we consider the points earned by riders in these competitions as individual projects.

Because the RCS assigns collaborative credit according to the performance of teammates within a team, mid-season transfers create an attribution problem. We therefore evaluate collaborative contribution separately within each team, so that a rider's value is linked to the teammates with whom the rider competed.\footnote{In Appendix \ref{app_riders}, we describe how we addressed the few riders who competed with two teams during the 2023 season.}

We compile the dataset using publicly available data from FirstCycling.com and ProCyclingStats.com. These databases reproduce official UCI results and rankings and provide detailed participation records that are not available in the official rankings. As shown in Table \ref{tab_data_A}, the final dataset contains 646 riders competing in 182 races, representing 36944 rider-days and 304835 UCI points. On average, a team participated in 86.8 races, ranging from 49 to 119, and each rider competed for an average of 57.5 days.

\begin{sidewaystable}[]
    \caption{Data summary}
    \label{tab_data_A}
\begin{tabular}{llcccccccc}
\toprule
{Numb} &{Team}  &{Name} &{Cat} &{Rk UCI} &{UCI pts} &{PCS pts} &{Riders} &{Races} & {Days}  \\ 
\midrule
1       & AG2R Citro{\"e}n Team        & AG2  & WT   & 18       & 9413       & 6386       & 30     & 93    & 1765 \\
2       & Alpecin-Deceuninck       & ALP  & WT   & 8        & 14900      & 8123       & 30     & 95    & 1667 \\
3       & Astana Qazaqstan Team    & AST  & WT   & 20       & 7123       & 4082       & 30     & 78    & 1725 \\
4       & Bahrain Victorious       & BAH  & WT   & 6        & 16122      & 9031       & 29     & 49    & 1521 \\
5       & BORA-hansgrohe           & BOR  & WT   & 10       & 13958      & 9118       & 30     & 79    & 1875 \\
6       & Cofidis                  & COF  & WT   & 15       & 10805      & 7065       & 30     & 118   & 1909 \\
7       & EF Education-EasyPost    & EF   & WT   & 11       & 12696      & 7694       & 30     & 76    & 1633 \\
8       & Groupama-FDJ             & GRU  & WT   & 7        & 15088      & 8673       & 28     & 81    & 1690 \\
9       & INEOS Grenadiers         & ING  & WT   & 4        & 18857      & 10708      & 30     & 62    & 1678 \\
10     & Intermarch\'e-Circus-Wanty & ITM  & WT   & 14       & 11228      & 7332       & 29     & 116   & 1802 \\
11     & Jumbo-Visma              & TJV  & WT   & 2        & 30047      & 14633      & 29     & 60    & 1521 \\
12     & Lidl-Trek                & LTK  & WT   & 5        & 16541      & 8697       & 29     & 82    & 1706 \\
13     & Movistar Team            & MOV  & WT   & 12       & 11285      & 7380       & 30     & 73    & 1765 \\
14     & Soudal Quick-Step        & SOU  & WT   & 3        & 19435      & 11524      & 29     & 85    & 1690 \\
15     & Team Ark\'ea-Samsic        & ARK  & WT   & 19       & 7585       & 5619       & 31     & 119   & 1885 \\
16     & Team dsm-firmenich       & DSM  & WT   & 17       & 9256       & 6093       & 30     & 79    & 1645 \\
17     & Team Jayco-AlUla         & JAY  & WT   & 13       & 11282      & 7135       & 30     & 84    & 1664 \\
18     & UAE Team Emirates        & UAE  & WT   & 1        & 32045      & 17169      & 30     & 92    & 1862 \\
19     & I. Premier Tech          & ISR  & PT   & 16       & 10227      & 6086       & 30     & 98    & 1504 \\
20     & Lotto Dstny              & LOT  & PT   & 9        & 14411      & 8796       & 29     & 117   & 1665 \\
21     & TotalEnergies            & TEG  & PT   & 22       & 5905       & 4164       & 25     & 93    & 1440 \\
22     & Uno-X Pro Cycling Team   & UNX  & PT   & 21       & 6626       & 4653       & 28     & 81    & 1332 \\
\midrule
\multicolumn{2}{l}{{Total}}           & & & & 304835     & 180161     & 646    & 1910  & 36944  \\
\multicolumn{2}{l}{{Average Team}}  &   &   &   & 13856.13   & 8189.14    & 29.36  & 86.82 & 1679.27 \\
\multicolumn{2}{l}{{Average Rider}} &   &   &   & 474.08     & 280.19     &        &       & 57.46  \\ 
\bottomrule
\end{tabular}
\end{sidewaystable}


\section{Results}
\label{sect_results}


\subsection{Distributional implications of alternative RCS configurations}
\label{subsect_results_rider_team}

\begin{figure}
  \centering
    \subfloat[Riders ranked 1st to 7th]{\label{fig_spider_a}$
        \includegraphics[height=44mm]{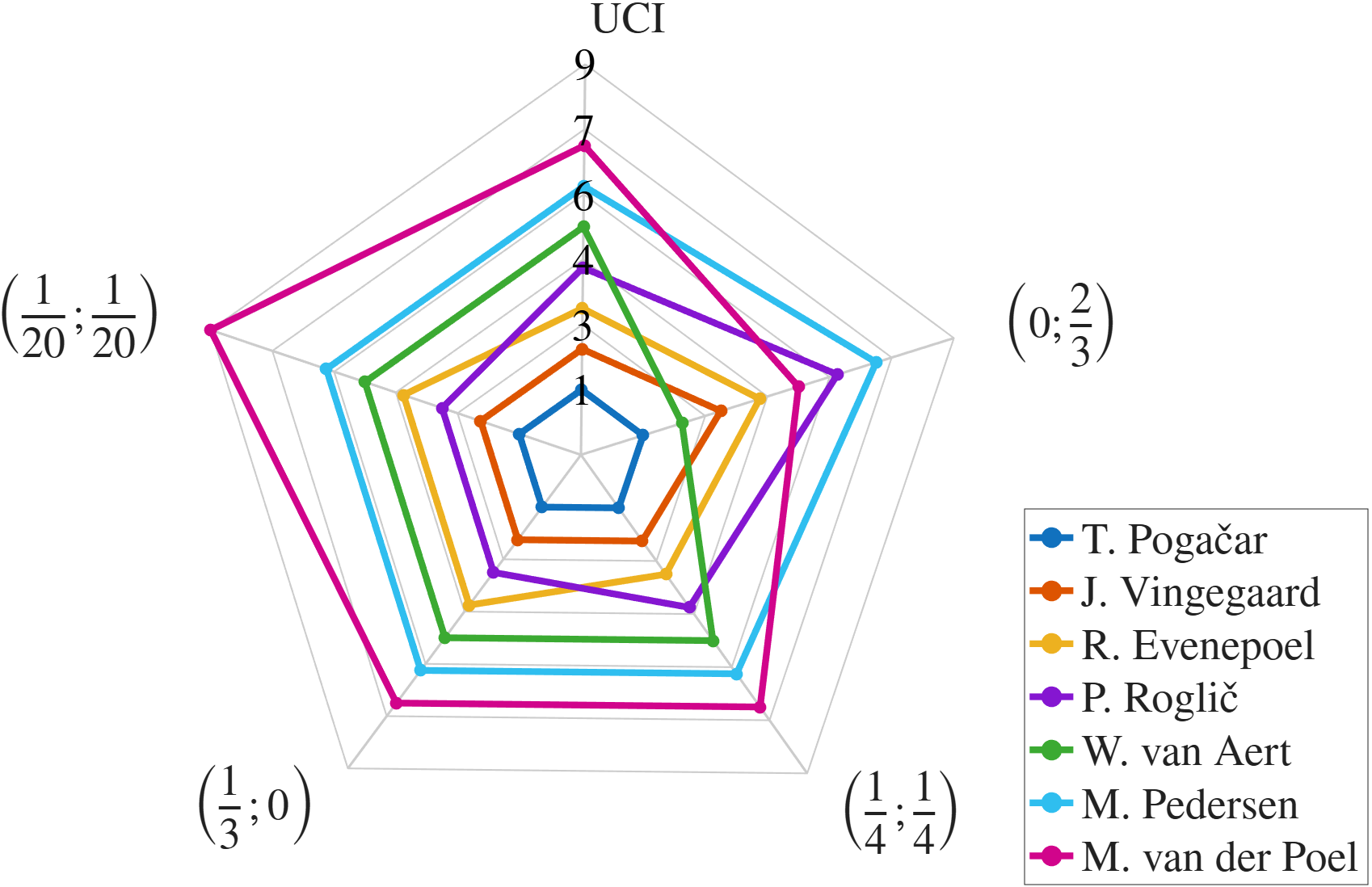}
      $}
      \hspace{0.10cm}
    \subfloat[Riders ranked 8th to 14th]{\label{fig_spider_b}$
        \includegraphics[height=44mm]{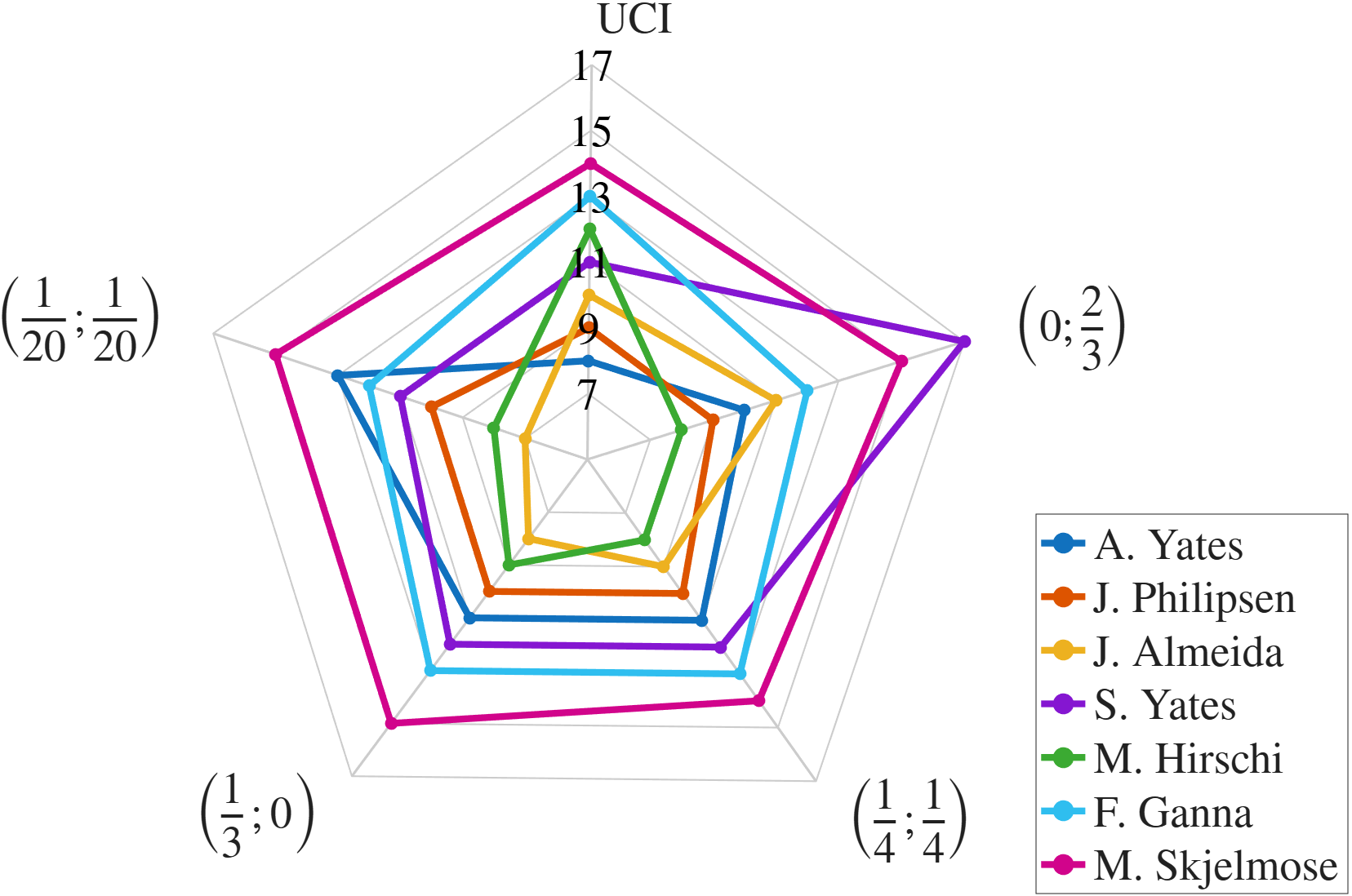}
      $}
    \caption{Rankings with equation \eqref{eq_dw_CoS} for riders ranked 1st to 14th with UCI points}
    \label{fig_spider}
  \end{figure}

Figure \ref{fig_spider} shows the rankings of the top 14 riders in the 2023 UCI ranking under various configurations of the parameters $\alpha$ and $\beta$ in equation \eqref{eq_dw_CoS}. These results reveal several notable patterns in the rider contribution assessment.

Regarding powerful riders who specialize in one-day races, both Wout van Aert and Mathieu van der Poel are among the top seven riders in the UCI ranking. They are consistently selected for the most prestigious races and move up in the ranking when the weight assigned to the participation component increases ($\alpha=0$ and $\beta=2/3$). The gap between the two riders widens when a greater weight is assigned to CoScore ($\alpha=1/20$ and $\beta=1/20$). van Aert keeps his 5th place due to the high number of points he earned and his contribution to his teammates' success in races such as the Tour de France and the Tirreno-Adriatico. In contrast, van der Poel drops in the ranking because he receives less redistributed credit from teammates than other top riders do. This is because his team, Alpecin-Deceuninck, primarily accumulated UCI points with three riders (see the discussion below).

In terms of general classification (GC) riders, Tadej Poga\v{c}ar and Jonas Vingegaard dominate all the rankings. The only change occurs under the participation specification ($\alpha=0$ and $\beta=2/3$), in which Vingegaard falls behind van Aert for the aforementioned reasons. Poga\v{c}ar retains the top position under this specification, suggesting that his performance combines strong results in stage races with substantial success in one-day races. Regarding other GC riders, as the weight assigned to CoScore increases (i.e., as $\alpha+\beta$ decreases), Remco Evenepoel and Primo\v{z} Rogli\v{c} swap positions in the ranking. This shift occurs because Evenepoel participated in 6834 points of his team's total, whereas Rogli\v{c} participated in 10154 points of his team's total, giving the latter more opportunities to benefit from his teammates' performance. 

Among the riders ranked from 8th to 14th (see Figure \ref{fig_spider_b}), three are Poga\v{c}ar's teammates at UAE Team Emirates. Adam Yates scored the most UCI points, but Jo\~{a}o Almeida and Marc Hirschi outperform him as the weight assigned to CoScore increases. This is because Yates scored fewer points than the other team's top riders in seven of the nine races in which they competed together. Almeida exhibits the greatest gain in points because he accumulated a substantial number of points on his own, and outperformed Yates and Hirschi in the three races in which they competed together. Hirschi likewise earned many UCI points independently. When competing alongside the team's other top riders, he had mixed results. However, he was the only one of the three riders who was able to outperform Poga\v{c}ar in some races.

This rider-level analysis illustrates a key limitation of CoScore. Sepp Kuss ranked 16th in the 2023 UCI standings and 11th in the participation configuration. However, his ranking drops to 138th when $(\alpha,\beta)=(1/20;1/20)$ is considered. In the three races where he was a domestique, his team won (Volta a Catalunya, Giro d'Italia, and Tour de France). In contrast, he only finished fifth in the UAE Tour when competing without the other team leaders. Furthermore, in his biggest victory, the Vuelta a Espa\~{n}a, Kuss narrowly outperformed Vingegaard and Rogli\v{c} by 162 and 470 points, respectively. In their previous races together, however, Vingegaard and Rogli\v{c} outscored Kuss by 2365 and 2120 points, respectively. Because CoScore is partly based on relative performance among teammates, a rider who repeatedly contributes to team victories while scoring substantially fewer points than the team's leaders may receive a low collaborative score. This motivates combining CoScore with individual performance and participation.

\begin{figure}
    \centering
      \includegraphics[width=140mm]{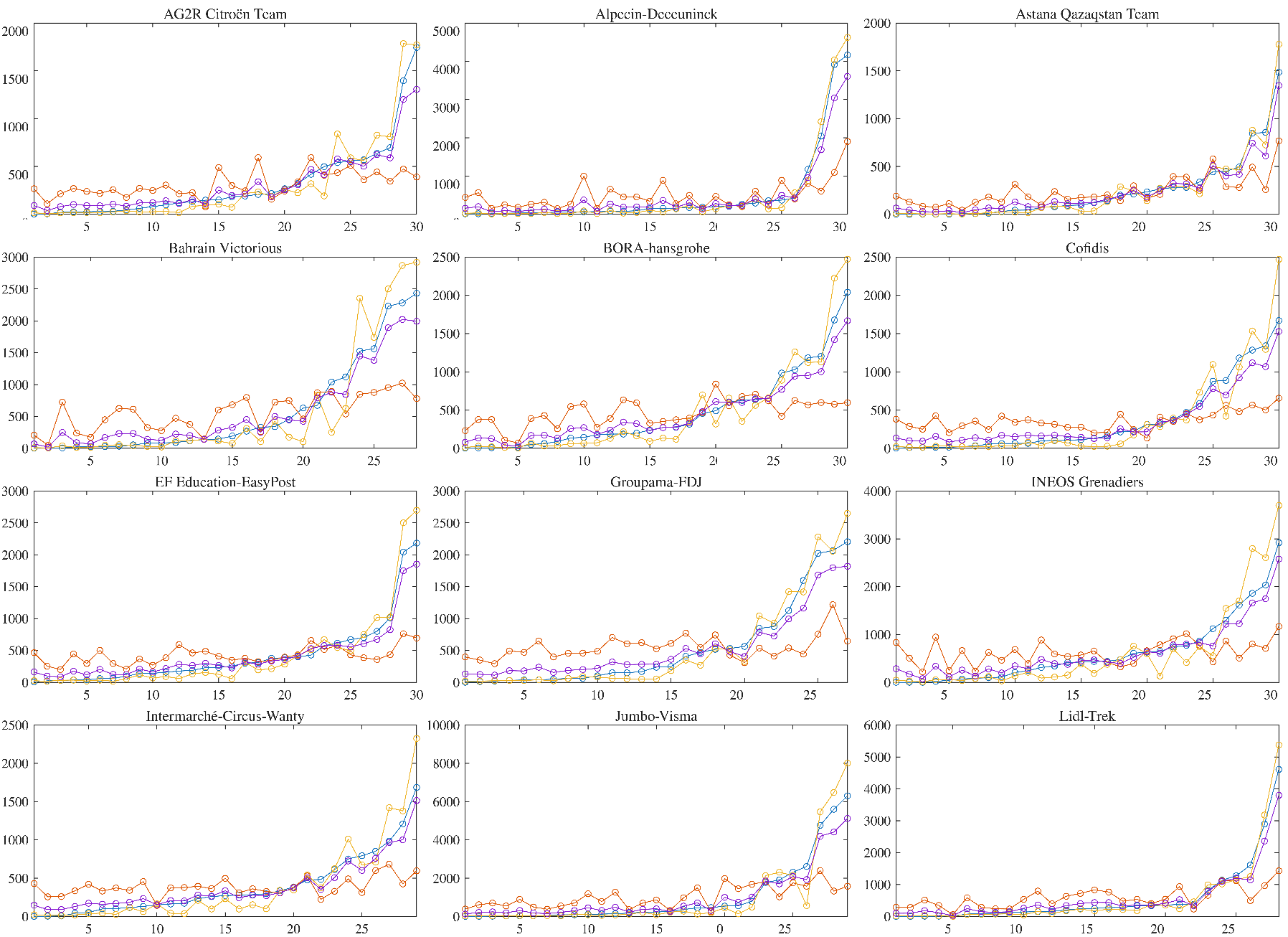}
      \includegraphics[width=140mm]{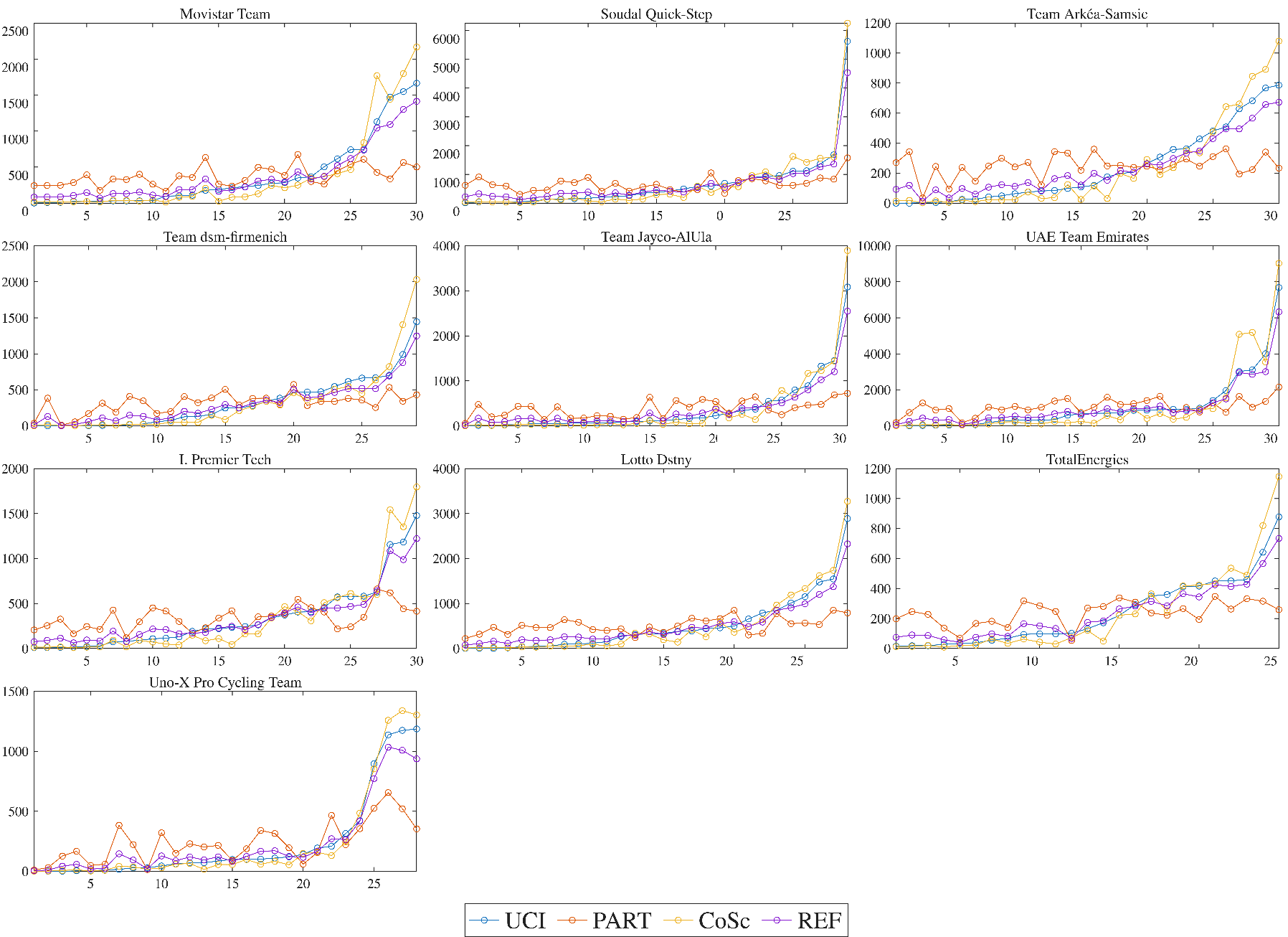}
    \caption{Distribution of rider productivity scores under UCI, PART, CoSc, and REF}
    \label{fig_lines}
  \end{figure}

Next, we examine how the distribution of UCI points changes within teams. Figure \ref{fig_lines} compares the allocation of each team's points under four illustrative configurations of equation \eqref{eq_dw_CoS}. First, we use the standard UCI points system, which is characterized by the parameters ($\alpha;\beta)=(1;0)$. Second, we consider the participation-weighted redistribution of points (PART), represented by ($\alpha;\beta)=(0;1)$. Third, we analyze the weighted CoScore value (CoSc), which is defined by ($\alpha;\beta)=(1/20;1/20)$. Finally, we consider a reference benchmark (REF) which is defined as the equally weighted version of equation \eqref{eq_dw_CoS} and is characterized by ($\alpha;\beta)=(1/3;1/3)$. To illustrate the different approaches adopted by teams during the season, we examine these point distributions by group in some squads into different categories.

In the first category, we can include teams characterized by concentrated point production. A notable example of this strategy is Alpecin-Deceuninck, which accumulated most of its points with three riders: Mathieu van der Poel, Jasper Philipsen, and Kaden Groves (67.9\% of the team's points). Lidl-Trek and Soudal Quick-Step also follows this kind of pattern. For these teams, Figure \ref{fig_lines} displays a very steep initial decline in the distribution of UCI points, followed by a long flat tail. The CoSc and REF specifications of the RCS exhibit a similar pattern. This pattern is consistent with a roster structure in which a relatively small number of elite riders account for most of the team's points.

The second category consists of the two dominant teams, UAE Team Emirates (UAE) and Jumbo-Visma (TJV). These two teams placed nine riders in the top 20 of the 2023 UCI ranking. Of all the riders on these teams, only Tiesj Benoot (TJV) increases his score under all three alternative configurations. This rider displays a remarkable balance between supporting team leaders and scoring points on his own. As for the riders who did not reach 1000 points, Diego Ulissi (UAE) increases his value with the PART configuration while maintaining his score under the CoSc configuration. This is because he scored a respectable number of points both when assisting his leaders and when competing independently.

We also identify a category of teams with balanced rosters. Bahrain Victorious and INEOS Grenadiers finished the season with seven and six riders scoring more than 1000 points, respectively. Consequently, mid-ranked riders in these teams tend to experience a reduction in their collaborative value due to competing alongside so many high scorers. An exception is Jack Haig, whose value increases with the REF specification. This is because in the races where he obtained most of his points he was able to outscore the team's leaders.

Finally, we identify a category of teams whose strategy is to accumulate a steady flow of UCI points throughout the season to secure a WT license. For Ark\'ea-Samsic, Team dsm-firmenich, and TotalEnergies the configurations of the RCS that we have considered produce rankings that are consistent with the UCI system. This shows that these teams achieve their results with their most valuable riders. In contrast, the rankings differ more substantially for Cofidis and Intermarch\'e-Circus-Wanty, suggesting the existence of valuable riders whose contributions are not fully reflected in the UCI ranking.

After analyzing changes at the rider and team levels, we use Bland-Altman plots to describe the systematic differences between the UCI system and the alternative RCS configurations. The plots display the difference between paired configurations against their average, with the conventional limits of agreement defined as the mean difference $\pm$1.96 standard deviations.

  \begin{figure}[t]
  \centering
    \subfloat[PART configuration]{\label{fig_ba_plot_part}$
        \includegraphics[height=53.5mm]{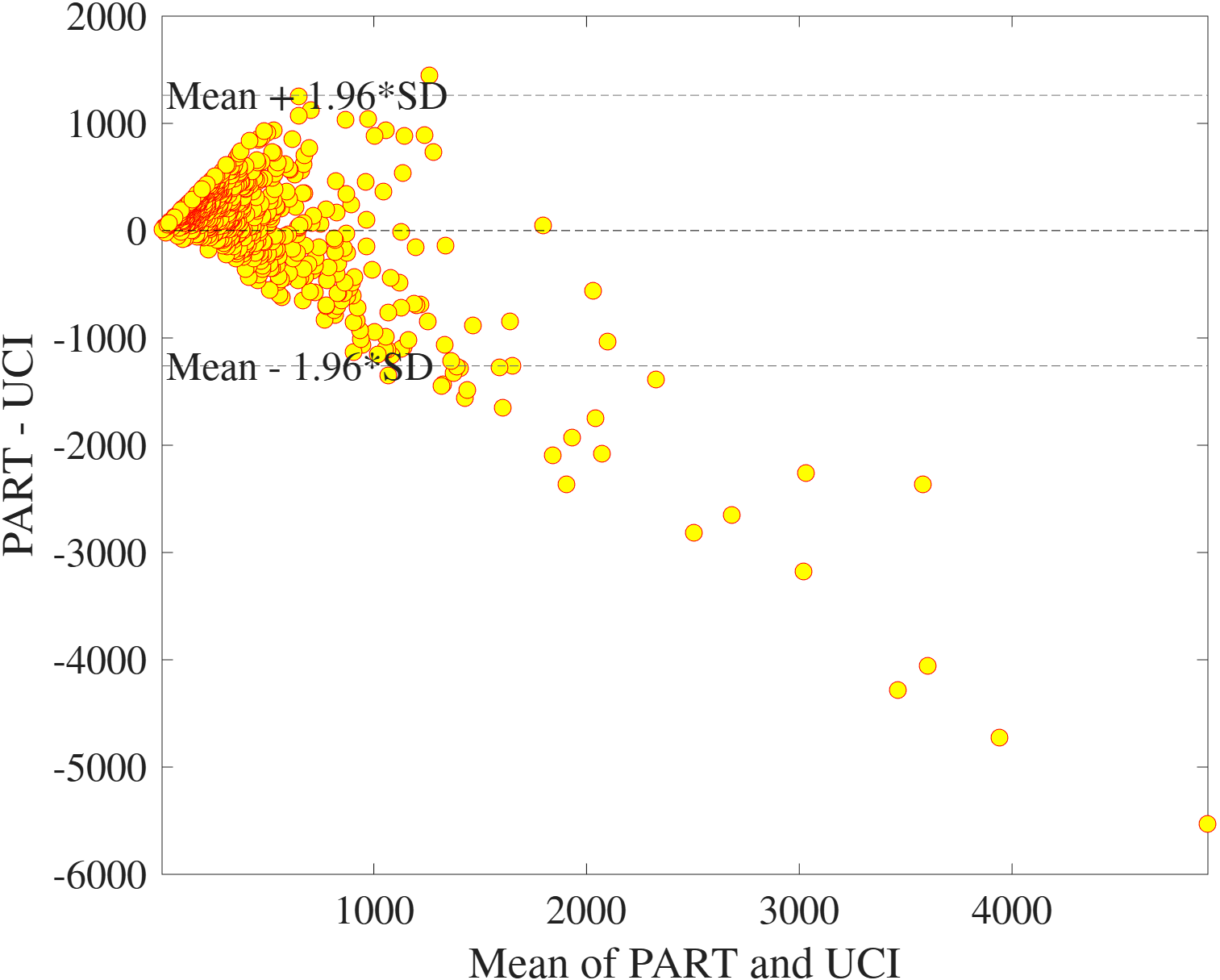}
      $}
      \hspace{0.15cm}
    \subfloat[CoSc configuration]{\label{fig_ba_plot_cosc}$
        \includegraphics[height=53.5mm]{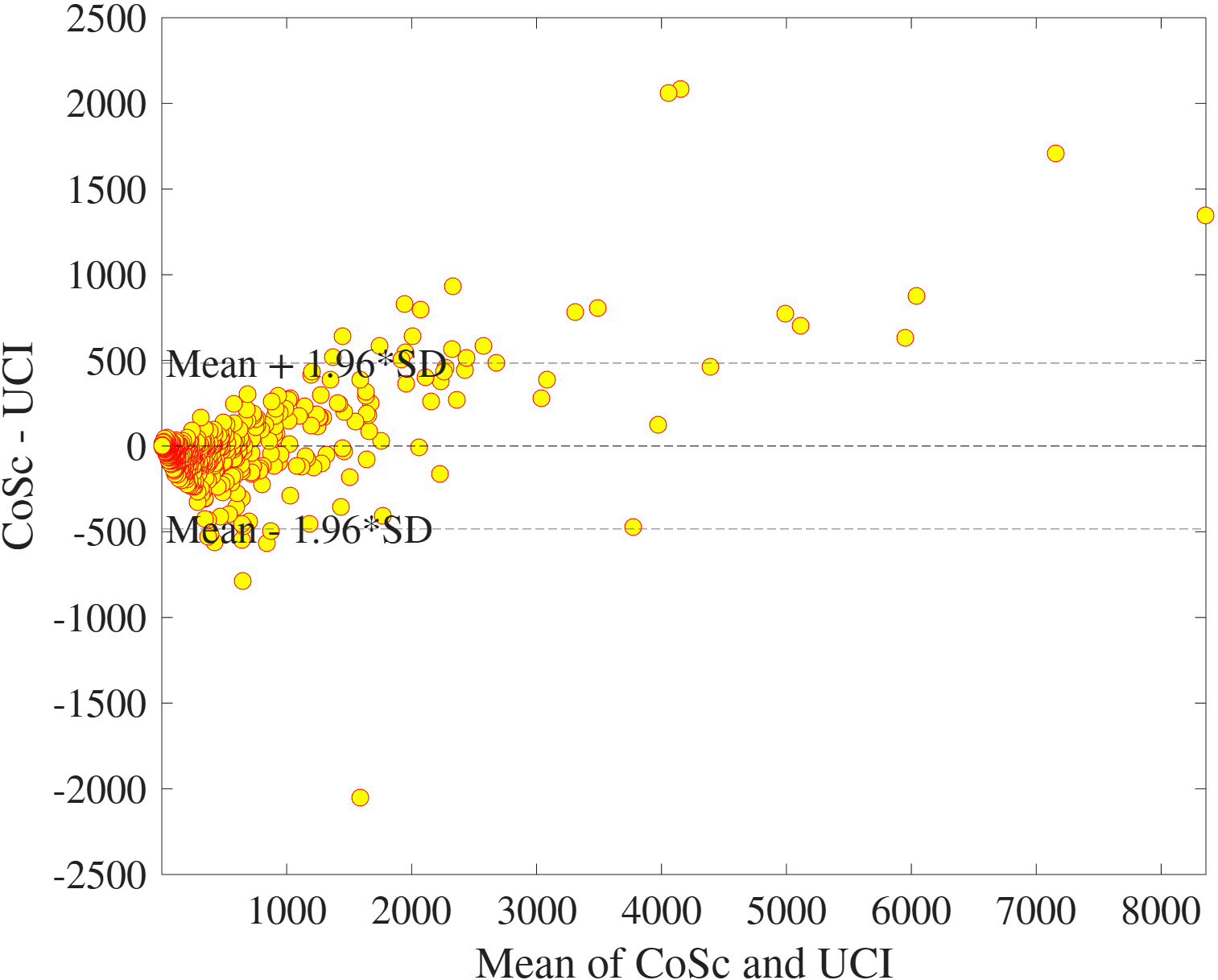}
      $}
      \vspace{0.45cm}
    \subfloat[REF configuration]{\label{fig_ba_plot_ref}$
        \includegraphics[height=53.5mm]{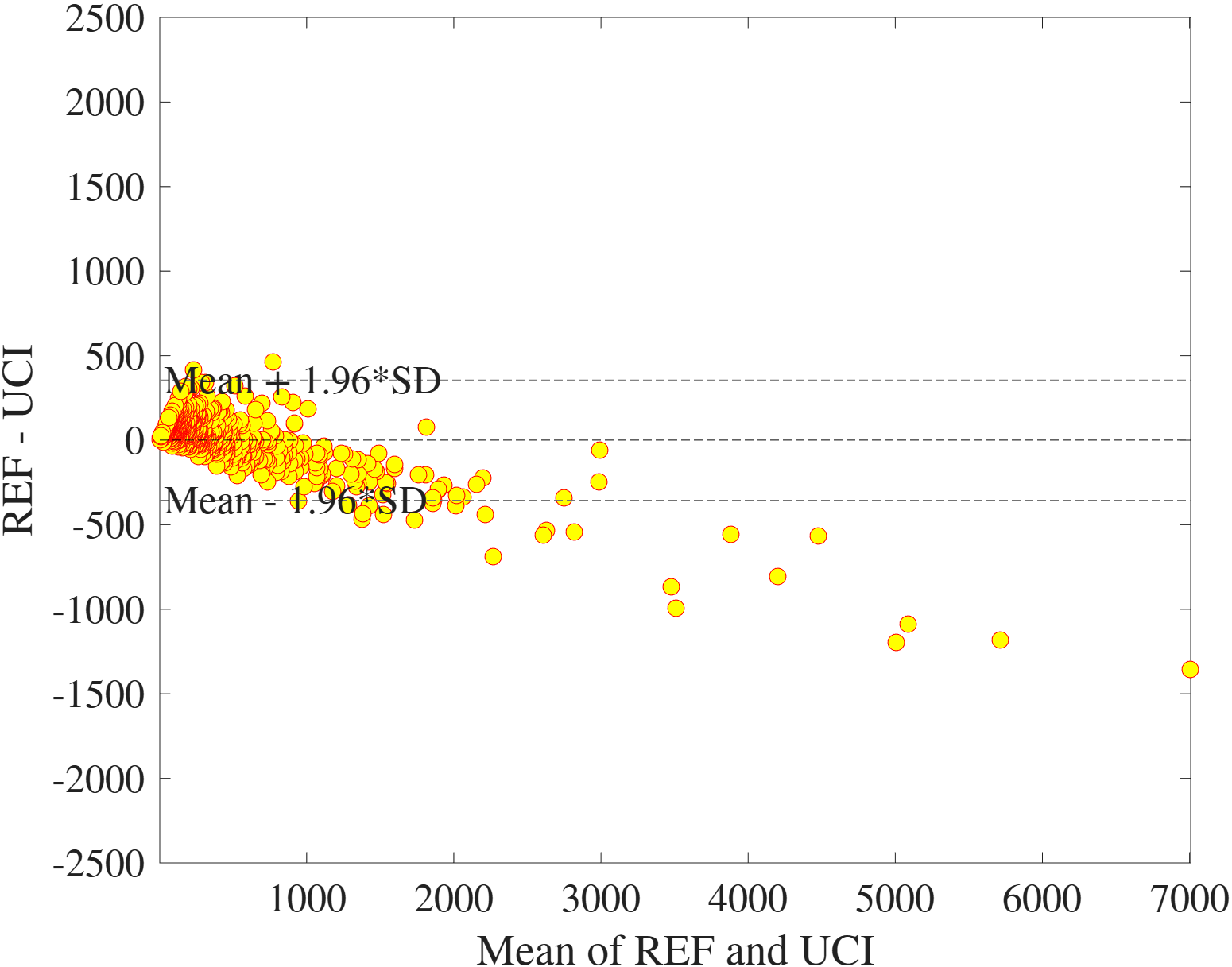}
      $}
    \caption{Bland-Altman plots between UCI and alternative configurations}
    \label{fig_ba_plot}
  \end{figure}

As Figure \ref{fig_ba_plot_part} shows, there is a significant disagreement between the UCI and the PART scores. This disagreement is quite symmetrical among those with an average score below 1000. Beyond this threshold, riders clearly receive fewer points with the PART specification than with the UCI system. This shows the redistributive feature of the PART configuration. In contrast, Figures \ref{fig_ba_plot_cosc} and \ref{fig_ba_plot_ref} show that the CoSc and the REF scores have a stronger agreement with UCI points. They also exhibit clear systematic differences in this agreement as the average number of points increases. Interestingly, these differences point in opposite directions. While the difference between the CoSc and the UCI scores increases with the average value, the difference between the REF and the UCI scores decreases. That is, unlike CoScore, the REF configuration produces a distribution of points from team leaders to domestiques.

Taken together, these results show that the RCS can substantially alter the allocation of team output. The direction and magnitude of the change depend on the weights assigned to the three performance indicators. Having established the redistributive properties of the framework, in Subsection \ref{subsect_parameters} we use an external benchmark to identify an interpretable reference specification for the RCS.


\subsection{Parameter calibration and reference specification}
\label{subsect_parameters}

The previously presented measure of collaborative contribution intentionally does not prescribe a unique evaluation philosophy. Instead, it gives managers, scouts, and analysts involved in talent identification and team strategy some discretion in identifying the characteristics they seek in a rider.

In this subsection, we use rider ratings from the Pro Cycling Manager (PCM) simulation game as an external benchmark to identify configurations that provide a plausible representation of rider evaluation. PCM is one of the few datasets that aggregates expert-informed assessments of riders across multiple dimensions. It is important to note that PCM's ratings reflect a consensus among data analysts and the cycling community rather than an objective measure of rider productivity. Moreover, since the ratings assess rider potential rather than team contribution directly, the benchmark should be interpreted as a test of external plausibility rather than the solution for the RCS.

PCM rates hundreds of real cyclists on a scale from 10 to 60. It also rates fourteen specific attributes ranging from 50 to 85 that reflect riders' skills in areas such as mountain climbing, sprinting, tactics, and recovery. Although PCM ratings may be subject to reputational bias, they provide an independent and generally accepted evaluation of a rider's potential. For our analysis, we create a PCM-based ranking using the average of the fourteen riding attributes. We use the ratings available at the beginning of the 2024 season so that the benchmark incorporates information available right after the 2023 season. Table \ref{tab_pcm} summarizes the data. There are 45 missing values for riders who retired after the 2023 season. 

\begin{table}[]
\caption{PCM individual values at the start of the 2024 season}\label{tab_pcm}
\begin{tabular*}{\textwidth}{@{\extracolsep\fill} lcccccc}
\toprule%
  Measure  & Observations & Min & Max & Mean & Median  & Variance \\
\midrule
   Potential    & 595  & 10     & 60     & 38.55  & 35    & 9.31   \\
   Attributes   & 595  & 60.14  & 78.79  & 68.63  & 68.43 & 2.89   \\
\bottomrule
\end{tabular*}
\end{table}

Because both PCM ratings and the RCS generate ordinal rankings, we use Kendall's $\tau$ to assess their rank-order association \citep[see][]{Kendall_38_Biom}. Table \ref{tab_kendall} reports the rank correlation coefficients between the PCM ranking and the rankings implied by different configurations of the RCS. Because the redistribution generated by the RCS may differ across the rider-performance distribution, we report correlations for the full sample (FS), the two halves of the distribution (H1 and H2), and the four quartiles (Q1, Q2, Q3, and Q4).

\begin{table}[h]
  \small
\caption{Kendall-$\tau$ rank correlation coefficients between PCM ratings and different collaborative contribution evaluations}\label{tab_kendall}
\begin{tabular*}{\textwidth}{@{\extracolsep\fill} lccccccc}
\toprule%
$(\alpha;\beta)$  & FS  & H1 & H2 & Q1 & Q2 & Q3 & Q4 \\
\midrule
$(1;0)$         & $0.512^{***}$ & $0.425^{***}$ & $0.265^{***}$ & $0.383^{***}$ & $0.146^{***}$ & $0.036$ & $0.241^{***}$ \\
$(2/3;0)$      & $0.500^{***}$ & $0.412^{***}$ & $0.257^{***}$ & $0.375^{***}$ & $0.137^{**}$   & $0.033$ & $0.240^{***}$ \\
$(1/2;0)$      & $0.492^{***}$ & $0.403^{***}$ & $0.252^{***}$ & $0.365^{***}$ & $0.131^{**}$   & $0.034$ & $0.236^{***}$ \\
$(1/3;0)$      & $0.480^{***}$ & $0.392^{***}$ & $0.243^{***}$ & $0.357^{***}$ & $0.128^{**}$   & $0.036$ & $0.222^{***}$ \\
$(1/10;0)$    & $0.454^{***}$ & $0.366^{***}$ & $0.230^{***}$ & $0.339^{***}$ & $0.110^{**}$   & $0.037$ & $0.203^{***}$ \\
\midrule
$(1/2;1/2)$       & $0.535^{***}$ & $0.416^{***}$ & $0.256^{***}$ & $0.400^{***}$ & $0.104^{*}$   & $0.031$ & $0.240^{***}$ \\
$(1/3;1/3)$    & $0.531^{***}$ & $0.413^{***}$ & $0.259^{***}$ & $0.386^{***}$ & $0.113^{**}$  & $0.032$ & $0.239^{***}$ \\
$(1/4;1/4)$    & $0.523^{***}$ & $0.408^{***}$ & $0.257^{***}$ & $0.379^{***}$ & $0.113^{**}$  & $0.034$ & $0.238^{***}$ \\
$(1/6;1/6)$    & $0.509^{***}$ & $0.398^{***}$ & $0.251^{***}$ & $0.372^{***}$ & $0.113^{**}$  & $0.041$ & $0.228^{***}$ \\
$(1/20;1/20)$  & $0.471^{***}$ & $0.367^{***}$ & $0.241^{***}$ & $0.343^{***}$ & $0.099^{*}$   & $0.046$ & $0.218^{***}$ \\
\midrule
$(0;1)$          & $0.391^{***}$ & $0.256^{***}$ & $0.177^{***}$ & $0.294^{***}$ & $0.014$        & $0.023$ & $0.194^{***}$ \\
$(0;2/3)$       & $0.491^{***}$ & $0.372^{***}$ & $0.220^{***}$ & $0.377^{***}$ & $0.077$        & $0.022$ & $0.215^{***}$ \\
$(0;1/2)$       & $0.508^{***}$ & $0.391^{***}$ & $0.229^{***}$ & $0.386^{***}$ & $0.093^{*}$   & $0.022$ & $0.218^{***}$ \\
$(0;1/3)$       & $0.508^{***}$ & $0.390^{***}$ & $0.233^{***}$ & $0.376^{***}$ & $0.099^{*}$   & $0.029$ & $0.214^{***}$ \\
$(0;1/10)$     & $0.478^{***}$ & $0.369^{***}$ & $0.235^{***}$ & $0.347^{***}$ & $0.098^{*}$   & $0.046$ & $0.212^{***}$ \\
\bottomrule
    \multicolumn{8}{l}{\begin{small}Significance level ($^{***}$) 1\%; ($^{**}$) 5\% and ($^{*}$) 10\%. \end{small}} 
\end{tabular*}
\end{table}

As shown in the table, the results are statistically significant for all subsamples except the third quartile. The smallest correlation coefficients are associated with $\alpha=0$, suggesting that PCM ratings are consistent with measures that incorporate race results. Similarly, the highest coefficients are found in the $\beta=0$ and $\alpha=\beta$ configurations, indicating that PCM assessments include UCI points. Notably, most of the top three coefficients are associated with the parameter region $\alpha=\beta$. This suggests that PCM ratings do not solely rely on official points. 

The results do not identify a uniquely optimal parameter configuration. The full-sample correlation is highest for the $(\alpha;\beta)=(1/2;1/2)$ configuration, while REF performs almost identically. The only configurations that consistently appear among the top five correlation coefficients across all subsamples are: UCI, REF, and $(\alpha;\beta)=(1/4;1/4)$. These results suggest that parametric configurations that balance individual performance and participation, while retaining a meaningful role for collaborative contribution, perform well against PCM assessments. Therefore, we adopt the equally weighted configuration $(\alpha=1/3$ and $\beta=1/3)$ as our reference specification. In Section \ref{subsect_regression}, we use this specification to examine the characteristics of riders associated with gains and losses under collaborative evaluation.


\subsection{Rider-level gains and losses under the RCS}
\label{subsect_regression}

We now examine which rider characteristics are systematically associated with the redistribution of points generated by the REF parametric configuration of the RCS. We consider two complementary dependent variables. The first one is the log of the ratio between the REF and the UCI specifications. Formally,
  \begin{equation*}
    \Delta^{P}_{i}=\log\left(\frac{REF_{i}+1}{UCI_{i}+1}\right).
  \end{equation*}

Defined in this way, the dependent variable accounts for the fact that a given change in points is relatively more important for a rider with a low initial number of points than for a rider at the top of the ranking.\footnote{A constant value of 1 is added to both measures to accommodate riders with zero UCI or REF points.} Furthermore, this specification makes the regression coefficients easier to interpret. Positive values of the dependent variable indicate that a rider gains points under the REF configuration compared to the UCI system, while negative values indicate that a rider loses points.

The second dependent variable is the change in a rider's ranking between the UCI and REF classifications. Positive values indicate an improvement in a rider's ranking with the REF configuration, while negative values indicate a deterioration. Formally,
  \begin{equation*}
    \Delta^{R}_{i} = Rank^{UCI}_{i} - Rank^{REF}_{i}.
  \end{equation*}

We use ordinary least squares (OLS) to estimate a series of linear regression models to examine the relationships between each dependent variable and two sets of explanatory variables (see Table \ref{tab_regression_variables} in Appendix \ref{app_add_tables} for a complete description of these variables). First, we consider a baseline model including the rider's age (\textit{Age}), the number of race days in which he participated (\textit{Days}), the Herfindahl-Hirschman index of UCI points concentration within the rider's team (\textit{TeamHHI}), his team's performance in UCI points per rider (\textit{TeamStrength}), and the log of the rider's UCI points plus 1 (\textit{LogUCI}). Subsequently, to check whether redistribution depends on rider specialization, we estimate an alternative model that incorporates interactions between \textit{LogUCI} and variables identifying different rider specializations (general classification, climbing, time trial, hilly terrain, sprinting, and one-day race). For each rider $i$ and each dependent variable $j=\{P,R\}$, we estimate the following two models: \\

\noindent\textbf{Model 1:}
\begin{equation*}
 \begin{split}
    \Delta^{j}_{i} & = \beta_{0} + \beta_{1}Age_{i} + \beta_{2}Days_{i} + \beta_{3}TeamHHI_{i} + \beta_{4}TeamStrength_{i}  \\
    & + \beta_{5} LogUCI_{i} + \varepsilon_{i}.
  \end{split}
\label{reg_model_1}
\end{equation*}

\noindent\textbf{Model 2:}
\begin{equation*}
 \begin{split}
    \Delta^{j}_{i} & = \beta_{0} + \beta_{1}Age_{i} + \beta_{2}Days_{i} + \beta_{3}TeamHHI_{i} + \beta_{4}TeamStrength_{i}  \\
    & + \beta_{5} LogUCI_{i} + \beta_{6}(LogUCI_{i} \times GC_{i}) + \beta_{7}(LogUCI_{i} \times Climb_{i}) \\
    & + \beta_{8}(LogUCI_{i} \times TT_{i}) + \beta_{9}(LogUCI_{i} \times Hill_{i}) \\
    & + \beta_{10}(LogUCI_{i} \times Sprint_{i}) + \varepsilon_{i}.
  \end{split}
\label{reg_model_2}
\end{equation*}

Since observations within the same team are not independent, statistical inference is based on cluster-robust standard errors with team-level clustering. Specifically, we use the CR2 small-sample correction and Satterthwaite-adjusted degrees of freedom for hypothesis testing. This approach is particularly appropriate given the relatively small number of clusters (teams), as conventional cluster-robust inference may otherwise yield overly precise estimates. All reported tests are two-sided, and statistical significance is assessed using Satterthwaite-adjusted $p$-values.

\begin{table}[htbp]
\centering
\caption{Regression results}
\label{tab_regression_results}
\begin{tabular}{lcccc}
\toprule
Variable & \multicolumn{2}{c}{Points} & \multicolumn{2}{c}{Ranks} \\
\cmidrule(lr){2-3} \cmidrule(lr){4-5} & Model (1) & Model (2) & Model (1) & Model (2) \\
\midrule
Intercept
& 2.15984*** & 2.21901*** & -95.805*** & -93.196*** \\
& (0.153402) & (0.145923) & (20.6881) & (19.4394) \\
Age
& 0.00453 & 0.00366 & 0.265 & 0.275 \\
& (0.003783) & (0.003722) & (0.5598) & (0.5357) \\
Days
& 0.01328*** & 0.01361*** & 1.916*** & 1.931*** \\
& (0.001757) & (0.001660) & (0.2269) & (0.2144) \\
TeamHHI
& -0.95817 & -1.11725 & -33.330 & -69.010 \\
& (0.574325) & (0.714842) & (85.7230) & (95.0876) \\
TeamStrength
& 0.00112*** & 0.00114*** & 0.168*** & 0.169*** \\
& (0.000165) & (0.000178) & (0.0263) & (0.0264) \\
LogUCI
& -0.57322*** & -0.56742*** & -19.443*** & -17.077*** \\
& (0.014370) & (0.014155) & (2.2253) & (2.4270) \\
\addlinespace
LogUCI$\times$GC 
& & -0.03822*** & & -5.562*** \\
& & (0.009286) & & (1.2811) \\
LogUCI$\times$Climb
& & 0.00882 & & -2.915** \\
& & (0.007579) & & (1.0706) \\
LogUCI$\times$TT
& & -0.01832** & & -0.194 \\
& & (0.006441) & & (1.2515) \\
LogUCI$\times$Hill
& & -0.02696*** & & -2.227* \\
& & (0.007816) & & (1.2383) \\
LogUCI$\times$Sprint
& & -0.01831* & & -4.152*** \\
& & (0.009420) & & (1.2978) \\
\bottomrule
\multicolumn{5}{l}{\footnotesize CR2 cluster-robust standard errors in parentheses.} \\
\multicolumn{5}{l}{\footnotesize Satterthwaite-adjusted degrees of freedom and $p$-values.} \\
\multicolumn{5}{l}{\footnotesize * $p<0.10$, ** $p<0.05$, *** $p<0.01$.}
\end{tabular}
\end{table}

Columns 2 and 4 of Table \ref{tab_regression_results} report the estimates from the baseline specification (Model 1) for the two dependent variables.\footnote{The complete estimation results are reported in Tables \ref{tab_regression_results_pts_mod_1} and \ref{tab_regression_results_pts_mod_2} in Appendix \ref{app_add_tables}.} The variable \textit{Days} is positively associated with redistribution under the REF specification. This may reflect the fact that participating in more races provides riders with more opportunities to contribute to the team's success. Holding all other variables constant, each additional race day is associated with a 1.34\% increase\footnote{The percentage increase in the dependent variable is calculated as $100(e^{\beta}-1)$.} in the $(REF + 1)/(UCI + 1)$ ratio and with an improvement of 1.92 ranking positions under REF relative to UCI. Age and the concentration index are not statistically significant. However, team performance has a positive and significant effect. Stronger teams tend to redistribute more collaborative value: an increase of 100 UCI points in the average team performance is associated with a 11.21\% increase in the $(REF + 1)/(UCI + 1)$ ratio and with an improvement of 16.8 ranking positions under REF relative to UCI. As discussed above, as riders' UCI performance increases, their REF allocation becomes progressively smaller relative to their UCI allocation. A 1\% increase in $UCI+1$ is associated with an approximately 0.57\% decrease in the REF-to-UCI ratio, holding the other variables constant. Similarly, riders with higher UCI scores tend to experience larger decreases in ranking under REF relative to UCI. Specifically, a 1\% increase in $UCI+1$ is associated with a reduction of about 0.19 ranking positions under REF allocation to UCI.

Columns 3 and 5 of Table \ref{tab_regression_results} report the estimates from the alternative specification (Model 2) for each dependent variable.\footnote{The complete estimation results are reported in Tables \ref{tab_regression_results_rks_mod_1} and \ref{tab_regression_results_rks_mod_2} in Appendix \ref{app_add_tables}.} This specification allows us to examine which types of riders benefit from or are disadvantaged by the collaborative evaluation. The estimated effects of \textit{Days}, \textit{TeamStrength}, and \textit{LogUCI} are similar to those in Model 1 in terms of value and statistical significance. This suggests that the alternative specification does not radically alter the basic relationships. Assuming one-day races as the reference rider specialization, the interaction terms exhibit a consistent pattern for \textit{GC}, \textit{Hill}, and \textit{Sprint}. All the coefficients are negative and statistically significant for both the point-change and rank-change models. For the former model, a 1\% increase in UCI points results in a larger decrease in the REF-to-UCI ratio for these three types of specialists than for those who specialized in one-day races. That is, the slope associated with \textit{LogUCI} is lower for \textit{GC}, \textit{Hill}, and \textit{Sprint} specialists than for the reference group. Similarly, for the rank-change model, a 1\% increase in UCI points generates larger decreases in the relative ranking for \textit{GC}, \textit{Hill}, and \textit{Sprint} than for one-day specialists. One possible explanation for these relationships is that GC and sprint specialists tend to rely more heavily on team support than one-day specialists. Indeed, GC riders and sprinters are paradigmatic examples of individuals who succeed with the help of their teammates. Specialists in hilly terrain tend to compete in races with many opportunities for breakaways, so they may need their teammates' help to control the race.

In contrast, interactions with \textit{Climb} and \textit{TT} specializations exhibit less consistent patterns. One possible explanation for the inconsistent results regarding climbers is that the PCS classification mixes riders of various profiles in this category. For example, it includes the most popular leader, Poga\v{c}ar, and the most popular domestique, Kuss. This makes it difficult to establish a clear relationship with respect to the reference specialization. Similarly, among time-trialists, there are riders who act as team leaders, such as Filippo Ganna, and riders who act as team helpers, such as Edoardo Affini.


\section{Discussion}
\label{sect_discussion}

UCI points determine race invitations and, consequently, sponsorship opportunities, so they strongly influence strategic behavior in professional road cycling. For example, in a stage of the Tour de France, 210, 150, 110, and 90 UCI points are awarded to the top four finishers, respectively. If a team has two riders in a four-rider breakaway, the team manager may instruct one of them to help their teammate compete for the 210-point victory. However, this could result in the domestique being caught by the peloton and losing their opportunity to obtain at least 90 points. This hypothetical situation illustrates the limitations of using rankings based on individual outcomes. These rankings may undervalue collaborative contribution and create incentives that differ from those associated with maximizing team success.

Redistributing resources among a given population is a central problem in the theory of fair allocation \citep[see][]{Thomson_19_Book}. The most popular mechanism for doing so is the proportional rule, which allocates resources in proportion to an individual characteristic. Considering a team's total UCI points as the resources and the set of days competed as the individual characteristic, the participation-weighted redistribution of points, represented by ($\alpha;\beta)=(0;1)$, applies this rule to the problem of a cycling season. CoScore, represented by ($\alpha;\beta)=(1/20;1/20)$, is a different redistribution mechanism that requires an individual's credit to depend not only on their own production record but also on their teammates' record in the project \citep[see][]{Flores-Szwagrzak_al_20_MS}. Finally, the UCI points system, represented by ($\alpha;\beta)=(1;0)$, can be understood as a sort of \textit{laissez-faire} distribution, in which each rider earns the points they obtained in the races in which they participated.

The study presented here shows that none of these three components captures all dimensions of rider contribution on its own. The official ranking undervalues domestique riders, the participation metric favors riders with extensive exposure to team success, and CoScore rewards riders who outperform their teammates.  

The alternative we considered here, the RCS, combines these three indicators of rider performance to create a flexible decision-support tool for evaluating collaborative contribution. Rather than identifying a single, universally optimal parameter configuration, the RCS should be selected according to the evaluator's objectives. Thus, it provides managers and analysts with a flexible tool for identifying talent, negotiating contracts, constructing rosters, and evaluating performance. In particular, the RCS offers a systematic approach to recognizing riders whose contributions to team success extend beyond their individual point accumulation. In brief, the PART configuration displays, by construction, a flat pattern that is useful for identifying riders who consistently compete in prestigious and successful events. In contrast, the distributions of points generated by CoSc and REF are similar to that of the UCI system, although they differ substantially from one another. As we mentioned before, CoSc produces a more concentrated distribution by assigning relatively more credit to riders who repeatedly outperform their high-scoring teammates, whereas REF compresses the distribution by balancing individual success with collaborative contribution.

Our study has some limitations. First, it relies on publicly available race results, so it does not account for tactical or contextual information during races. Second, the calibration exercise is based on subjective assessments rather than objective measures of rider value. Finally, we infer rider roles from observed performance, but we do not take into account that team responsibilities may evolve during the season. 

Future research could extend the framework using additional physical and tactical information, such as power output, physiological characteristics, and time spent in breakaways. Such data could improve both the identification of collaborative roles and the calibration of the framework's weighting parameters.

Although our method for allocating collective output was inspired by male professional cycling, it could be applied to other categories where data is scarce, such as women's professional cycling or lower competitive levels, if calibrated correctly. Additionally, it would be interesting to explore the applicability of the RCS to other collaborative sports where individual contributions must be evaluated within collective production processes. 

Beyond professional sports, our framework illustrates a general challenge faced by organizational environments where there is a potential conflict between individual recognition and group effort. As \cite{Flores-Szwagrzak_al_20_MS} point out, academic researcher rankings face similar problems, especially when evaluating joint scientific production by senior and junior researchers. Similarly, assessing the performance of environmental, economic, or technological policies within supranational entities presents similar challenges when some countries have stronger backgrounds than others.


\section{Conclusions}
\label{sect_conclusions}

Evaluating individual contributions in collaborative environments is challenging because observable outcomes often reflect the combined efforts of several participants rather than a single individual's performance. Professional road cycling provides a particularly suitable setting to study this problem, as team success depends on the coordinated actions of riders performing complementary roles. 

This paper addresses this issue by proposing the Rider Contribution Score (RCS), a measure that combines three complementary dimensions of rider performance: UCI points, race participation, and CoScore. This measure provides analysts, managers and researchers  with an intuitive and easy-to-implement solution to the issue of undervaluing riders whose roles prioritize team outcomes over personal recognition.

Using data from the 2023 professional road cycling season, we have shown that the proposed framework offers a meaningful alternative perspective on rider evaluation compared to official UCI rankings. The RCS places greater value on riders who earn a significant number of points, especially when they do so without competing alongside their team's top scorers. The RCS also values riders who consistently support team leaders, provided the team achieves strong overall results and the supporting riders make meaningful individual contributions. In contrast, the RCS assigns relatively less value to high-scoring cyclists whose results are primarily obtained in races where the team performs exceptionally well. Furthermore, our empirical analysis shows that riders with high UCI point totals tend to be at a disadvantage with the REF specification, especially for GC, sprint, and hill terrain specialization.


\section*{Disclosure statement}

\subsection*{Ethical Approval}

Not applicable.

\subsection*{Competing interests}

The author declares no competing interests.

\subsection*{Authors' contributions}

Not applicable.

\subsection*{Funding}

Financial support from grant PID2023-146364NB-I00, funded by MCIU/AEI/10.13039/501100011033 and FSE+.

\subsection*{Availability of data and materials}

Data from cycling races is publicly available at firstcycling.com and procyclingstats.com. PCM data can be downloaded from pcmdaily.com.


\vspace{0.50cm}

\appendix
\section*{Appendix}
\addcontentsline{toc}{section}{Appendix}

\def\thesubsection{A.\arabic{subsection}}
\linespread{1.15}


\subsection{Riders competing with two teams}
\label{app_riders}

  \begin{enumerate}
    \item Trainees and riders from development squads: we have excluded them from the sample because the UCI accumulates their points to their usual teams. Example: Simon Dehairs (Alpecin-Deceuninck).
    \item Riders promoted to one of the 22 teams in our sample: we have only considered the points and days raced after the promotion took place. There is one case: Alec Segaert (Lotto Dstny).
    \item Riders who participated in UCI-sanctioned races with their national teams: we have excluded their results in these events. Examples: Caleb Ewan (Australia) and Simone Velasco (Italy).
    \item Mid-season transfers: in cases where only one team is included in the sample, we have removed the results from the other team. There are two cases: Jan Christen (UAE Team Emirates) and Andrii Ponomar (Team Ark\'{e}a-Samsic). If both teams are included in the sample, each rider is treated as two separate observations to reflect the actual productivity of their teammates. Next, we have excluded these riders from the ranking analysis. There are three cases: Arnaud D\'{e}mare (Groupama-FDJ, transferred to Team Ark\'{e}a-Samsic), Harm Vanhoucke (Team dsm-firmenich, transferred to Lotto Dstny), and Antonio Tiberi (Bahrain Victorious, after his contract was terminated by Lidl-Trek).
  \end{enumerate}
  

\subsection{Tables}
\label{app_add_tables}

\begin{sidewaystable}[h]
\centering
\caption{Explanatory variables for the differences between the REF and UCI evaluations}
\label{tab_regression_variables}
\begin{tabular}{p{2.55cm} p{12.40cm} c c c}
\toprule
\textbf{Variable} & \textbf{Description} & \textbf{Min} & \textbf{Max} & \textbf{Average} \\
\midrule
Age & Age, in years, of rider $i$ at the end of the 2023 season
& 19 & 41 & 28.12 \\
Days & Total number of race days started by rider $i$ during the 2023 season
& 2 & 89 & 57.5 \\
TeamHHI & Herfindahl-Hirschman concentration index of the distribution of UCI points within rider $i$'s team during the 2023 season
& 0.066 & 0.176 & 0.097 \\
TeamStrength & Average UCI points per rider in rider $i$'s team during the 2023 season
& 236.2 & 1068.2 & 472.1 \\
LogUCI & Log of rider $i$'s UCI points during the 2023 season plus 1
& 0 & 8.947 & 5.037 \\
OneDay & Indicator equal to one if rider $i$'s specialization according to \textit{ProCyclingStats} is one-day races, and zero otherwise
& 0 & 1 & 0.323 \\
GC & Indicator equal to one if rider $i$'s specialization according to \textit{ProCyclingStats} is general classification, and zero otherwise
& 0 & 1 & 0.278 \\
TT & Indicator equal to one if rider $i$'s specialization according to \textit{ProCyclingStats} is time trial, and zero otherwise
& 0 & 1 & 0.109 \\
Sprint & Indicator equal to one if rider $i$'s specialization according to \textit{ProCyclingStats} is sprint, and zero otherwise
& 0 & 1 & 0.095 \\
Climb & Indicator equal to one if rider $i$'s specialization according to \textit{ProCyclingStats} is climb, and zero otherwise
& 0 & 1 & 0.108 \\
Hill & Indicator equal to one if rider $i$'s specialization according to \textit{ProCyclingStats} is hill, and zero otherwise
& 0 & 1 & 0.086 \\
\bottomrule
\end{tabular}
\end{sidewaystable}

\begin{table}[h]
\centering
\caption{Regression results: Points Model (1)}
\label{tab_regression_results_pts_mod_1}
\begin{tabular}{lccccc}
\toprule
Variable & Estimate & SE & $t$-stat. & df (Satt.) & $p$-value \\
\midrule
Intercept & 2.15984*** & 0.153402 & 14.08 & 18.20 & $<0.001$ \\
Age & 0.00453 & 0.003783 & 1.20 & 18.37 & 0.24648 \\
Days & 0.01328*** & 0.001757 & 7.56 & 18.96 & $<0.001$ \\
TeamHHI & -0.95817 & 0.574325 & -1.67 & 3.70 & 0.17638 \\
TeamStrength & 0.00112*** & 0.000165 & 6.79 & 3.93 & 0.00262 \\
LogUCI & -0.57322*** & 0.014370 & -39.89 & 19.36 & $<0.001$ \\
\bottomrule
\multicolumn{6}{l}{\footnotesize Multiple R-squared: 0.845, Adjusted R-squared: 0.8438. CR2 cluster-robust standard} \\
\multicolumn{6}{l}{\footnotesize errors. Satterthwaite-adjusted degrees of freedom and $p$-values.} \\
\multicolumn{6}{l}{\footnotesize * $p<0.10$, ** $p<0.05$, *** $p<0.01$.}
\end{tabular}
\end{table}

\begin{table}[h]
\centering
\caption{Regression results: Points Model (2)}
\label{tab_regression_results_pts_mod_2}
\begin{tabular}{lccccc}
\toprule
Variable & Estimate & SE & $t$-stat. & df (Satt.) & $p$-value \\
\midrule
Intercept & 2.21901*** & 0.145923 & 15.207 & 18.32 & $<0.001$ \\
Age & 0.00366 & 0.003722 & 0.984 & 18.33 & 0.33777 \\
Days & 0.01361*** & 0.001660 & 8.196 & 19.13 & $<0.001$ \\
TeamHHI & -1.11725 & 0.714842 & -1.563 & 3.84 & 0.19605 \\
TeamStrength & 0.00114*** & 0.000178 & 6.376 & 3.99 & 0.00314 \\
LogUCI & -0.56742*** & 0.014155 & -40.085 & 19.19 & $<0.001$ \\
\addlinespace
LogUCI$\times$GC & -0.03822*** & 0.009286 & -4.116 & 20.00 & $<0.001$ \\
LogUCI$\times$Climb & 0.00882 & 0.007579 & 1.164 & 17.33 & 0.26015 \\
LogUCI$\times$TT & -0.01832** & 0.006441 & -2.845 & 15.38 & 0.01206 \\
LogUCI$\times$Hill & -0.02696*** & 0.007816 & -3.449 & 17.46 & 0.00297 \\
LogUCI$\times$Sprint & -0.01831* & 0.009420 & -1.944 & 18.94 & 0.06687 \\
\bottomrule
\multicolumn{6}{l}{\footnotesize Multiple R-squared: 0.8528, Adjusted R-squared: 0.8504. CR2 cluster-robust standard} \\
\multicolumn{6}{l}{\footnotesize errors. Satterthwaite-adjusted degrees of freedom and $p$-values.} \\
\multicolumn{6}{l}{\footnotesize * $p<0.10$, ** $p<0.05$, *** $p<0.01$.}
\end{tabular}
\end{table}

\begin{table}[h]
\centering
\caption{Regression results: Ranks Model (1)}
\label{tab_regression_results_rks_mod_1}
\begin{tabular}{lccccc}
\toprule
Variable & Estimate & SE & $t$-stat. & df (Satt.) & $p$-value \\
\midrule
Intercept & -95.805*** & 20.6881 & -4.631 & 18.20 & $<0.001$ \\
Age & 0.265 & 0.5598 & 0.473 & 18.37 & 0.64206 \\
Days & 1.916*** & 0.2269 & 8.441 & 18.96 & $<0.001$ \\
TeamHHI & -33.330 & 85.7230 & -0.389 & 3.70 & 0.71875 \\
TeamStrength & 0.168*** & 0.0263 & 6.404 & 3.93 & 0.00326 \\
LogUCI & -19.443*** & 2.2253 & -8.737 & 19.36 & $<0.001$ \\
\bottomrule
\multicolumn{6}{l}{\footnotesize Multiple R-squared: 0.4582, Adjusted R-squared: 0.4539. CR2 cluster-robust standard} \\
\multicolumn{6}{l}{\footnotesize errors. Satterthwaite-adjusted degrees of freedom and $p$-values.} \\
\multicolumn{6}{l}{\footnotesize * $p<0.10$, ** $p<0.05$, *** $p<0.01$.}
\end{tabular}
\end{table}

\begin{table}[h]
\centering
\caption{Regression results: Ranks Model (2)}
\label{tab_regression_results_rks_mod_2}
\begin{tabular}{lccccc}
\toprule
Variable & Estimate & SE & $t$-stat. & df (Satt.) & $p$-value \\
\midrule
Intercept & -93.196*** & 19.4394 & -4.794 & 18.32 & $<0.001$ \\
Age & 0.275 & 0.5357 & 0.514 & 18.33 & 0.61339 \\
Days & 1.931*** & 0.2144 & 9.005 & 19.13 & $<0.001$ \\
TeamHHI & -69.010 & 95.0876 & -0.726 & 3.84 & 0.50976 \\
TeamStrength & 0.169*** & 0.0264 & 6.390 & 3.99 & 0.00312 \\
LogUCI & -17.077*** & 2.4270 & -7.036 & 19.19 & $<0.001$ \\
LogUCI$\times$GC & -5.562*** & 1.2811 & -4.341 & 20.00 & $<0.001$ \\
LogUCI$\times$Climb & -2.915** & 1.0706 & -2.723 & 17.33 & 0.01429 \\
LogUCI$\times$TT & -0.194 & 1.2515 & -0.155 & 15.38 & 0.87855 \\
LogUCI$\times$Hill & -2.227* & 1.2383 & -1.798 & 17.46 & 0.08946 \\
LogUCI$\times$Sprint & -4.152*** & 1.2978 & -3.199 & 18.94 & 0.00474 \\
\bottomrule
\multicolumn{6}{l}{\footnotesize Multiple R-squared: 0.4876, Adjusted R-squared: 0.4795. CR2 cluster-robust standard} \\
\multicolumn{6}{l}{\footnotesize errors. Satterthwaite-adjusted degrees of freedom and $p$-values.} \\
\multicolumn{6}{l}{\footnotesize * $p<0.10$, ** $p<0.05$, *** $p<0.01$.}
\end{tabular}
\end{table}


\bibliographystyle{chicago}
\bibliography{references}


\end{document}